\documentstyle[12pt,epsfig]{article}
\newlength{\dinwidth}
\newlength{\dinmargin}
\newcommand\ben{\begin{equation}}
\newcommand\een{\end{equation}}
\newcommand\bea{\begin{eqnarray}}
\newcommand\eea{\end{eqnarray}}

\newcommand\Tr{{\rm Tr}}
\newcommand\nn{\nonumber}

\newcommand\del{\partial}

\newcommand\tg{\tilde{g}}

\newcommand{\beq}{\begin{equation}}
\newcommand{\eeq}{\end{equation}}
\newcommand{\beqa}{\begin{eqnarray}}
\newcommand{\eeqa}{\end{eqnarray}}
\newcommand{\be}{\begin{equation}}
\newcommand{\ee}{\end{equation}}
\newcommand{\bc}{\begin{center}}
\newcommand{\ec}{\end{center}}
\newcommand{\al}{\alpha}

\newcommand{\ra}{\rightarrow}

\newcommand{\ie}{{\em i.e.}}
\newcommand{\eg}{{\em e.g.}}

\newcommand\tX{{\tilde{X}}}

\begin{document}
\thispagestyle{empty}
\addtocounter{page}{-1}
\vskip-0.35cm
\begin{flushright}
UK/08-11 \\
TIFR-TH/08-28 \\
\end{flushright}
\vspace*{0.2cm}
\centerline{\Large \bf   Gauge Theories with Time Dependent Couplings }
\centerline{\Large \bf and their Cosmological Duals}
\vspace*{1.0cm}
\centerline{\bf Adel Awad${}^{a,b}$\footnote{On leave of absence from
Ain Shams University, Cairo, EGYPT}, Sumit R. Das${}^a$, Suresh Nampuri${}^d$,}
\centerline{\bf K. Narayan ${}^c$ and Sandip P. Trivedi ${}^d$}
\vspace*{0.2cm}
\centerline{\it Department of Physics and Astronomy,}
\vspace*{0.2cm}
\centerline{\it University of Kentucky, Lexington, KY 40506 \rm USA ${}^a$}
\vspace*{0.35cm}
\centerline{\it Center for Theoretical Physics, British University of
  Egypt}
\vspace*{0.2cm}
\centerline{\it Sherouk City 11837, P.O. Box 43, EGYPT ${}^b$}
\vspace*{0.35cm}
\centerline{\it Tata Institute of Fundamental Research,}
\vspace*{0.2cm}
\centerline{\it Mumbai 400005, \rm INDIA ${}^d$}
\vspace*{0.35cm}
\centerline{\it Chennai Mathematical Institute,}
\vspace*{0.2cm}
\centerline{\it Padur PO, Siruseri 603103, \rm INDIA ${}^c$}
\vspace*{1.0cm}
\centerline{\tt adel@pa.uky.edu, das@pa.uky.edu}
\centerline{\tt
nampuri@gmail.com, narayan@cmi.ac.in, sandip@tifr.res.in}

\vspace*{0.8cm} \centerline{\bf Abstract} \vspace*{0.3cm}

We consider the $N=4$ SYM theory in flat $3+1$ dimensional spacetime
with a time dependent coupling constant which vanishes at $t=0$, like
$g_{YM}^2=t^p$. In an analogous quantum mechanics toy model we find that
the response is singular. The energy diverges at $t=0$, for a generic state. 
 In addition, if $p>1$ the phase of the wave function has
a wildly oscillating behavior, which does not allow it to be continued
past $t=0$. A similar effect would make the gauge theory singular as well,
though nontrivial effects of renormalization could tame this singularity
and allow a smooth continuation beyond $t=0$.
The gravity dual in some cases is known to be a time
dependent cosmology which exhibits a space-like singularity at $t=0$.
Our results, if applicable in the gauge theory for the case of the
vanishing coupling, imply that the singularity is a genuine sickness
and does not admit a meaningful continuation.
When the coupling remains
non-zero and becomes small at $t=0$, the curvature in the bulk becomes
of order the string scale. The gauge theory now admits a time
evolution beyond this point.  In this case, a finite amount of energy
is produced which possibly thermalizes and leads to a black hole in
the bulk.

\vspace*{0.5cm}
\baselineskip=18pt

\tableofcontents

\newpage

\section{Introduction and Summary}
\label{sec:intro}

The resolution of singularities is an outstanding problem in the study
of gravity.  The gauge theory/gravity correspondence provides a
non-perturbative framework for the study of gravity, and one would
hope that it can shed some light on this question.

With this motivation some cosmological solutions which admit natural
gauge theory duals have been constructed and studied recently
\cite{DMNT1,DMNT2,ADNT,Chu:2006pa,Lin:2006ie}.  These solutions can be
thought of as deformations of $AdS_5\times S^5$ (in Poincare
coordinates) and are dual to strongly coupled ${\cal N}=4$ Yang Mills
theory in flat $3+1$ dimensional space-time subjected to external time
dependent or null sources. For other classes of solutions of this
type, see \cite{Chen:2005ae},\cite{ohta}. A related approach to gauge
theory duals of cosmological singularities has been pursued in
\cite{hertog} and more recently in \cite{Turok:2007ry}. Ideas about
finding signatures of spacelike singularities inside black holes in
the dual gauge theory are described in \cite{Kraus:2002iv}.

In this paper our interest is in the time dependent cases. These have
been further studied in \cite{ADNT}. Here, the only source is a time
dependent coupling of the gauge theory.  At early times, the 'tHooft
coupling in the gauge theory is large and varies slowly.  The AdS/CFT
correspondence tells us that the gravity dual is a non-normalizable
deformation of $AdS_5 \times S^5$ sourced by the dilaton field.  As
time evolves, the gauge theory state evolves in response to the time
dependent coupling. On the gravity side, the background evolves
according to the supergravity equation of motion, subject to
appropriate boundary conditions.  In particular the dilaton starts
back reacting leading to a nontrivial metric.

Our main interest will be in situations in which the dilaton,
 $e^\Phi$, vanishes at time $t=0$. In the corresponding gravity
 solution a space-like singularity appears at $t=0$ which extends all
 the way to the boundary \footnote{However, the $4$ dim. metric as
 seen in the gauge theory is flat, after a conformal transformation,
 as discussed in section ~\ref{sec:cosmosol}.}.  One would like to
 know if this singularity is a genuine sickness in the theory or if it
 merely signals a breakdown of the supergravity approximation.  Since
 the bulk theory has a boundary dual which is formulated in a precise
 fashion, we can ask this question in the dual theory. One would like
 to know if the boundary theory is sick at the singularity or if it
 allows for a continuation past the point where the dilaton vanishes
 \footnote{More precisely, the supergravity approximation breaks down
 before the singularity forms, once the curvature gets to be of order
 the string scale. Using the boundary theory we would like to find out
 if there is a continuation past this region.}.

In this paper we try to analyze this question in some detail.
Prompted by the cosmological solutions we ask the following general
question first: Consider the $ {\cal N}=4$ Yang Mills theory subjected
to an external time dependent dilaton. We take the dilaton to be of
the form,
\be
\label{dilprof}
e^\Phi=(-t)^p, ~~~~~~~~~t\le 0,
\ee
so that it vanishes as $t\rightarrow 0^-$. In eq.(\ref{dilprof}), the
index $p$ can take any positive real value.  In such a situation the
question we ask is whether the response of the gauge theory to this
external source is singular or not, as $t\rightarrow 0^-$ ?

It is useful to first consider a quantum mechanics model which has an
analogous coupling to the dilaton. We find that the response in this
quantum mechanical system is singular, for all values of $p>0$, in the
following sense: The time dependent dilaton pumps energy into the
system, and as $t\rightarrow 0^-$ we find that the energy diverges.
The nature of the singularity does depend on the index $p$. For $p\ge
1$, when the variation of the dilaton is more rapid, the wave function
of the system acquires a time dependent phase factor which becomes
wildly oscillating and diverges as $t\rightarrow 0^-$.  As a result
the wave function of the system (in the Schrodinger picture) does not
have a well defined limit as $t\rightarrow 0^-$.  In contrast for
$p<1$ the phase factor does not diverge and the wave function has a
well defined limit as $t\rightarrow 0^-$. Even so, the energy diverges
as $t\rightarrow 0^-$ (this also happens when $p\ge 1$).

The result that for $p\ge 1$ the wave function becomes singular,
without a well defined limit, holds regardless of the state of the
system.  On the other hand, the conclusion that the energy diverges
is true for a generic state. For special states in which an
appropriate matrix element vanishes, the energy can remain finite, as
we discuss below \footnote{Fluctuations in the energy will diverge in
these special states, even if the expectation value of the energy
remains finite. However such fluctuations are suppressed at leading
order in $N$.}.

The analysis in the quantum mechanics model is most conveniently
carried out in the Schrodinger picture. The conclusions stated above
follow from the fact that near $t=0$ the potential energy term in the
Schrodinger equation dominates the time evolution and the Kinetic
energy term is subdominant.

One can carry out a similar analysis in the gauge theory. Once again
if we assume that the potential energy dominates near $t=0$ one finds
that the behaviour is completely analogous to that in the quantum
mechanics model disussed above.  The energy diverges as $t\rightarrow
0^-$ and for $p>1$ the wave function acquires a wildly oscillating
time dependent phase which does not have a well defined limit as $t
\rightarrow 0^-$.
 
However, in the field theory case, we have not been able to establish
that the approximation leading to the conclusions above definitely
holds.  Differences with the quantum mechanics model arise due to the
infinite number of modes in field theory. These have to be dealt with
carefully by regulating the theory after introducing a cutoff and
incorporating the effects of renormalisation in this cutoff theory. We
have not carried out this procedure adequately to determine whether
the approximation mentioned above of the potential energy dominating
holds, and our conclusions about the gauge theory response being
singular are therefore tentative and not definite. As we discuss
below, higher loop effects could sum up to render the kinetic term
important and tame the singularity, resulting in finite energy
production and a smooth continuation of the wavefunctional beyond
$t=0$.  Studying this further will require both a better understanding
of the calculational aspects of renormalisation in the time dependent
gauge theories at hand, and a better understanding of the conceptual
issues involved in incorporating these effects of renormalisation in
the Schrodinger picture which we use in this paper
\footnote{We thank David Gross and the referee for emphasising these
points to us.}.

The AdS cosmologies described in \cite{DMNT1,ADNT} correspond to the
value $p=\sqrt{3}$ in eq.(\ref{dilprof}).  From the discussion above
it follows that, if our approximation of a dominant potential energy
continues to hold in the gauge theory, the resulting singularity is a
genuine sickness which does not admit a well defined continuation.
Since $p>1$ in this case, the wave function in the gauge theory
description becomes wildly oscillating without a well defined limit as
$t\rightarrow 0^-$. Also the energy should diverge as $t\rightarrow
0^-$.

Our analysis shows that the singularity in the bulk arises  due to two
related reasons.  First, the dilaton vanishes, resulting in the string
frame curvature blowing up.  Second, an infinite amount of energy is
dumped into the system by the vanishing dilaton, resulting in a
singular back reaction.

We do not know the explicit form of the bulk solutions whose
boundaries are conformally flat for values of
$p$, other than $\sqrt{3}$.  However such solutions should exist since
we are specifying the dilaton field on the boundary for all times.  It
is worth pointing out that the singular behavior we find is not tied
per se to the non-analyticity of the dilaton as $t\rightarrow 0$, and
occurs for all values of $p>0$, integer and non-integer.  Rather, the
singular behavior is related to the rate at which the dilaton
vanishes.  As was mentioned above, with our approximations, the
behavior is more singular for $p\ge 1$, when the dilaton vanishes more
rapidly, than it is for $p<1$.

A few more comments about the analysis in the gauge theory are also
worth making.  At first sight one might think that when the dilaton
vanishes the gauge theory becomes weakly coupled and can be analyzed
in perturbation theory.  This turns out not to be true.  Starting from
a generic state, one finds that the time dependent dilaton excites the
fields to large enough values so that the cubic and quartic
interaction terms are non-negligible near the point where the dilaton
vanishes and perturbation theory is not valid. This is the essential
reason why the analysis gets complicated.  Based on the quantum
mechanics model we find that the Schrodinger picture is particularly
useful in analyzing the resulting behavior.  A WKB-like approximation
can be formulated in this picture in the vicinity of the vanishing
dilaton.  This allows the leading behavior near the singularity to be
analyzed without having to resort to perturbation theory. As was
mentioned above, in the gauge theory, we have not been able to
establish that this approximation is indeed correct and thus our
conclusions should be taken to be indicative rather than definitive.

It is worth emphasizing that if the potential energy term continues to
dominate near the singularity, our conclusions
about the gauge theory in the presence of a coupling which truly
vanishes are valid at finite $N$ and finite $g_{YM}^2N$ and are
therefore a result about the dual closed string theory in the presence
of stringy and/or quantum corrections. As we will see, the behavior of
the wavefunctional near the time of vanishing coupling is essentially
determined by the coupling constant and not the details of the
lagrangian.

It follows from our analysis that to find cosmological solutions which
 are not sick the dilaton's behavior at the boundary has to be
 modified so that it does not vanish.  Once this is done, for a
 smoothly varying dilaton, one does not expect the gauge theory to be
 singular \footnote{We are assuming here that the gauge theory does
 not have any phase transitions as the dilaton is varied. If this
 assumption is wrong the response of the gauge theory to a smoothly
 varying dilaton need not be non-singular.  Such a phase transition
 can be avoided by working on $S^3$, at finite $N$. We thank S. Wadia
 for emphasizing this point to us.}.  In a situation where the dilaton
 profile is chosen to be constant in the far future, reaching a value
 such that the 'tHooft coupling in the boundary theory is large, one
 expects that the supergravity solution becomes a black hole in the
 far future.  This is based on the expectation that the dual gauge
 theory will generically have some non-vanishing energy density in the
 far future and this energy will eventually thermalize.  It is worth
 noting that if this expectation is met, the spacetime which is highly
 curved when the dilaton is small, eventually evolves to a smoothly
 varying spacetime outside the black hole horizon. The fate of the
 theory will be similar if renormalization effects tame the gauge
 theory even in the case where the coupling truly vanishes.

 An important question
which we cannot address here is how the thermalization process
depends on the dilaton time profile. To answer this question one needs
a better understanding of the system when the 'tHooft coupling is of
order one, for example.

One might wonder if the formation of the black hole can be avoided in
a situation where the dilaton profile is time reversal invariant.  In
such a case states should exist which evolve in a time reversal
invariant manner.  Classically, in such a state all velocities have to
vanish at $t=0$; quantum mechanically, the wave function has to be
real at $t=0$.  Starting from $t=0$ in such a state and evolving into
the future one expects that a black hole will typically form, if some
net energy is input into the system. Thus we do not expect the absence
of a black hole in such states, rather these states will correspond to
starting with a black hole in the far past and ending with one in the
far future. In some situations, this conclusion might be avoided, but
we do not understand at the moment how to identify them \footnote{The
case where the boundary theory is on $S^3$ rather than $R^3$ is more
promising in this respect, since the formation of a black hole then
requires the temperature to be bigger than that of the Hawking Page
transition.}.

The breakdown of perturbation theory near $t=0$, in the time dependent
case, is quite different from what happens in the null-dependent
solutions which were studied in \cite{DMNT1,DMNT2,Chu:2006pa}.  In the
null case, the effects we describe in this paper are absent, there is
no particle production and perturbation theory is possibly applicable
for correlators of fields at light front times when the coupling is
small.

We discussed cosmological solutions above with $p=\sqrt{3}$, which
were studied in \cite{DMNT1,ADNT}.  These correspond to symmetric
Kasner-like solutions, where all spatial directions expand at the same
rate, or to FRW solutions.  There are other asymmetric Kasner
solutions which were also constructed in \cite{DMNT1,ADNT}.  The dual
gauge theory in these cases does not live in flat space but instead in
a space-time with a curvature singularity, which occurs at the time
coincident with the bulk singularity. Since the boundary metric is
non-dynamical, it is difficult to see how time evolution on the
boundary can be continued past the singularity in such circumstances
leading to the conclusions that the boundary hologram is sick in these
cases as well.

The behavior of the cosmological solutions near the singularity has
 some degree of universality.  For example in the symmetric Kasner and
 FRW solutions, mentioned above, the spatial curvature is different
 but this feature is irrelevant near the singularity where the
 evolution of the spacetime is determined by the diverging dilaton
 stress tensor.  One expects this to be a more general feature- some
 differences among solutions should become irrelevant leading to the
 same behavior near the singularity.  Our conclusions obtained from
 studying the gauge theory dual to the cosmological solutions, within
 our approximations, do not require any particular state, and apply
 quite generally.  One therefore expects these conclusions to hold for
 all solutions with the same common behavior near the singularity,
 e.g., for all conformally flat four dimensional metrics as discussed
 in section \ref{sec:bklcosmo}.

To explore this further we discuss Bianchi IX type cosmologies in the
 presence of the dilaton towards the end of this paper. These
 solutions are constructed by taking solutions of 4 dim. gravity
 coupled to dilaton and embedding them in $5$ dim. AdS space as
 discussed in \cite{DMNT1,ADNT}.  Interestingly, once the dilaton is
 excited only a finite number of BKL oscillations occur between Kasner
 regimes. With each oscillation the importance of the dilaton stress
 energy grows, at the expense of the spatial curvature. Eventually all
 solutions enter a Kasner regime, where all directions shrink, and the
 spatial curvature is relatively unimportant.  In a whole family of
 solutions this final Kasner regime corresponds to the symmetric
 Kasner solution with $p=\sqrt{3}$, mentioned above. For other
 solutions the final Kasner regime is one where the three spatial
 directions shrink in an asymmetric manner.  The holographic dual for
 these cosmologies can be constructed. Our analysis of the singularity
 in the symmetric and asymmetric Kasner solutions then applies to all
 these solutions.  It will be interesting to explore these solutions
 and their holographic duals further.

This paper is organized as follows. In section~\ref{sec:gtandm} we
discuss the gauge theory and introduce a toy field theory which
captures the essential features of the gauge theory in the dilaton
background. We also introduce a quantum mechanics model which has many
features in common with these field theories. The quantum mechanics
model is then analyzed in considerable detail in
section~\ref{sec:da}. The implications for the toy field theory and
the gauge theory are discussed in section~\ref{sec:gt}. The
connections to the cosmological solutions are discussed in
section~\ref{sec:cosmosol}. Other Kasner and BKL-like solutions are
discussed in section~\ref{sec:bklcosmo}.

Several important details are in the appendices.
Appendix A discusses the coupling of the dilaton to  the Yang Mills theory.
Appendix B discusses the time dependent Harmonic oscillator. Subleading
corrections to the energy, in the vicinity of the vanishing dilaton, are
 discussed in Appendix C. Particle production in  the quadratic approximation,
for a modified non-vanishing dilaton profile, is discussed in Appendix D.
The behavior of the Yang Mills theory in the presence of a  dilaton which
varies with Milne time is studied  in Appendix E.
Finally, some discussion about the universal behavior near singularities and
about  BKL cosmologies in the presence of the dilaton,
 is contained in Appendix F.

\section{The Gauge Theory and a Toy Model}\label{sec:gtandm}

We will consider the $N=4$ gauge theory defined on a flat $3+1$
dimensional space-time, which is regarded as the Poincare patch
boundary of an AdS cosmology - e.g. the ones discussed in detail in
\cite{ADNT}.
The bulk dilaton is equal to the coupling constant of the Yang Mills theory,
\be
\label{dil}
e^{\Phi(t)}=g_{YM}^2.
\ee
Thus the varying dilaton gives rise to a varying Yang Mills coupling.

After suitable field redefinitions
 the Lagrangian for the ${\cal N}=4$ theory
takes the form,
\begin{eqnarray}
\label{lagn4}
L& = & {\rm Tr} \{
-{1\over 4 e^{\Phi}}F_{\mu\nu}F^{\mu\nu} + {1\over 2} (D_\mu X^a)^2
-{1\over 4}e^{\Phi}([X^a,X^b])^2 \cr
&& -i{\bar \Psi }\Gamma^\mu D_\mu \Psi + e^{\Phi/2}
{\bar \Psi}\Gamma^a[X^a,\Psi] \}.
\end{eqnarray}
There are six scalars, $X^a, a=1, \cdots 6$. And $4$ two-component
Weyl fermions of $SO(1,3)$, which have been grouped together as one
Majorana Weyl Fermion of $SO(1,9)$. The Gamma matrices $\Gamma^\mu,
\mu=0,1, \cdots 3$, and $ \Gamma^a, a= 1, \cdots 6$, together form the
$10$ Gamma matrices of $SO(1,9)$.  The scalars and fermions transform
as the adjoint of $SU(N)$. The covariant derivative of the scalars is,
\be
\label{covderv}
D_\mu X^a=\partial_\mu-i[A_\mu,X^a],
\ee
and similarly for the fermionic fields. The field strength is,
\be
\label{fstrenght}
F_{\mu\nu}=\partial_\mu A_\nu-\partial_\nu A_\mu+i[A_\mu,A_\nu].
\ee

In appendix A we discuss   the Lagrangian above, especially the
dilaton couplings,  in more detail.
Let us make one comment here. It is sometimes stated that the dilaton
couples to the  Lagrangian of the ${\cal N}=4$ theory, \cite{wittene}.
 The more correct statement is that  the dilaton couples to the on-shell Lagrangian \cite{keni}.
This  differs from the operator obtained from  eq.(\ref{lagn4}) but only  by a total derivative
term involving the scalars.

At the singularity, $e^\Phi$ goes to zero. As a result, the prefactor in
the gauge kinetic energy term in eq.(\ref{lagn4}) blows up.  This is
the essential complication which must be dealt with in our analysis of
the time dependent situation at hand. To study it further it is
useful to introduce a toy model consisting of a single scalar field
$\tilde{X}$ with Lagrangian

\be
\label{toy}
L=-{1\over  e^{\Phi}}[{1\over 2} (\partial {\tilde X})^2  + {\tilde X}^4].
\ee

Note that the quartic term is ``right side up'',
so that for a constant dilaton this model would have a stable minimum.
We will see that this model captures all the essential features of the real problem of interest.

In fact it is useful to simplify the model further and to consider a quantum mechanics system, with action,
\be
\label{qmsys}
S=\int dt {1\over e^{\Phi}}[{1\over 2} \dot{\tilde{X}}^2-{1\over 2} \omega_0^2 \tilde{X}^2 -\tilde{X}^4]
\ee
Starting from the field theory eq.(\ref{toy}) such a system would arise if we keep only one fourier mode of momentum ${\vec k}$ with $\omega_0^2=k^2$, and only the quartic self interactions of this mode.

It will be useful in the discussion below to first consider the quantum mechanics
system and then return to the field theories eq.(\ref{toy}) and eq.(\ref{lagn4}).

\section{ Analysis of  Quantum Mechanics Model}\label{sec:da}
We now turn to analyzing the quantum mechanics  model eq.(\ref{qmsys}) further.

It is useful to begin  by carrying out a field redefinition which gives rise
to a variable  with a canonical kinetic energy term and  analyzing the
system in terms of this new variable. We define the variable $X$ to be
\be
\label{nvarx}
X=e^{-\Phi/2} {\tilde X}.
\ee
Upto a surface term the Lagrangian is
\be
\label{lagnew}
L= {1\over 2}\dot{X}^2 - {1\over 2} \omega^2(t) X^2 - e^{\Phi} X^4.
\ee
Here  $\omega^2(t)$ is  a time dependent angular frequency that arises due to the time dependent dilaton and is given by
\be
\label{mass}
\omega^2(t)=-[{1\over 4}({\dot \Phi})^2 - {1\over 2} {\ddot \Phi}]+\omega_0^2,
\ee
where the dot superscript  indicates a time derivative of the dilaton.
For a dilaton dependence
\be
\label{gendil}
e^\Phi=(-t)^p, t<0,
\ee
we have
\be
\label{masstwo}
\omega^2(t)= - {\alpha^2\over t^2} +\omega_0^2
\ee
with,
\be
\label{valalpha1}
\alpha^2={p\over4}(p+2).
\ee
Note that for sufficiently small time, $\omega^2$ becomes negative, and  the variable  $X$ has a tachyonic mass term (negative $(mass)^2$), which arises due to the
 time dependent dilaton.
 In fact $\omega^2$   diverges as $t\rightarrow 0$,
this will be important
in the subsequent analysis.

Let us briefly outline the detailed analysis that follows.
Since the $X^4$ term in eq.(\ref{lagnew}) is multiplied by $e^\Phi$, one might at first
expect that this term  is not important near $t=0$. Accordingly, we first neglect it
and analyze the resulting quadratic theory. It turns out that
 due to the diverging tachyonic mass the system quite generally
gets driven to $|X|\rightarrow \infty$ as
$t\rightarrow 0$. As a result, the quantum mechanical description in terms of the
$X$ variable is not complete. If we want to know what happens as $t\rightarrow 0$,
and beyond, one needs additional information about the behavior at
$X\rightarrow \infty$.  The diverging value of $X$ also means that the quartic term
cannot be neglected.

At this stage it is worth remembering that the field redefinition, eq.(\ref{nvarx}),
is singular at $t=0$. In fact, we find that the rate at which $X$ diverges is exactly
balanced by the rate at which the dilaton vanishes, leaving $\tilde{X}$ to be
finite, as $t\rightarrow 0$.
This motivates us to study the system in terms of  the $\tilde{X}$ variable.
Although, as mentioned above, the analysis in terms of the $X$ variable
has already revealed that  the
quartic term cannot be neglected,  we ignore it at first, to gain some understanding of
the system. Our analysis shows that the system is singular in a manner we have
described in the   section~\ref{sec:intro}.
 We then incorporate the quartic terms and find that all the
essential conclusions about the singular nature of the response go through unchanged.

\subsection{The $X$ Description }
As was mentioned above, to begin we drop the quartic term in eq.(\ref{lagnew}).
this gives rise to the quadratic action,
\be
\label{lagsmt}
S=\int dt{1\over 2}[\dot{X}^2-\omega(t)^2X^2]
\ee
The quadratic theory  can then  be analyzed in standard fashion by expanding the field,
$X$, in terms of normal modes,


From eq.(\ref{lagsmt})
it follows that the corresponding operator, $\hat{X}$, in the Heisenberg picture,
satisfies the equation
\be
\label{eqo}
\ddot{\hat{X}}+\omega^2(t)\hat{X}=0.
\ee

Let us define $f(t)$ to be,
\be
\label{deff}
f(t) = \sqrt{\pi \omega_0\over 2} \sqrt{-t}
H_\nu^{1}(-\omega_0t),
\ee
where,
 \be
\label{defnu}
\nu= {p+1\over 2},
\ee
and $H_\nu^{(1)}(x)$ is the Hankel function which asymptotically, as $x\rightarrow \infty$, behaves like
\be
\label{asshank}
H_\nu^{(1)}(x) \rightarrow \sqrt{2\over \pi x}e^{[i(x-(\nu+{1\over 2}){\pi \over 2})]}.
\ee

  Then it follows from the standard properties of Hankel functions that $f(t)$
  satisfies the equation,
\be
\label{eqf}
\ddot{f(t)}+\omega^2(t) f(t)=0,
\ee
with boundary condition,
\be
\label{asf}
f(t) \rightarrow e^{-i\omega_0t}, t\rightarrow -\infty.
\ee
The solution to eq.(\ref{eqo}) now is,
\be
\label{hatx}
\hat{X}={1\over\sqrt{2 \omega_0}}[{\hat a} f(t) + (\hat{a})^\dagger f^*(t)].
\ee
The momentum conjugate to $\hat{X}$ is
\be
\label{defp}
\hat{P}=\dot{\hat{X}}={1\over\sqrt{2 \omega_0}}[{\hat a} \dot{f}(t) + (\hat{a})^\dagger \dot{f}^*(t)].
\ee
The operators, $\hat{X}$, and $\hat{P},$ satisfy the canonical commutation relation iff,
 ${\hat a}, {\hat a}^\dagger,$ satisfy
the standard relation,
\be
\label{srel}
[\hat{a}, \hat{a}^\dagger]=1.
\ee

 Classical solutions to eq.(\ref{eqo}) take the form,
\be
\label{csol}
X(t)=\sqrt{(-t)}[ A J_\nu(-t) + B N_\nu(-t)],
\ee
where $J_\nu, N_\nu,$ stand for the Bessel and Neumann functions. The constants $A,B,$
 are determined by the
initial conditions, at $t\rightarrow -\infty$. For  generic initial conditions, $B\ne 0$, and  as
$t\rightarrow 0^-$,
\be
\label{bX}
X(t)\sim (-t)^{{1\over 2}-\nu} = (-t)^{({-p\over2})} \rightarrow \infty.
\ee
Here we have used eq.(\ref{defnu}), and the fact that $p>0$.
 Thus we see that due to the negative and diverging value of $\omega^2$ near the singularity, a generic
trajectory gets driven out to infinite values of the position coordinate.

Classical states correspond to coherent states in the quantum theory. We see that the center of
the wave packet for a generic coherent state runs away to infinity due to the tachyonic mass term.
This shows that further data is needed to make the quantum theory in terms of the variable $X$, and
 Lagrangian, eq.(\ref{lagsmt}),  well defined.  This additional data should specify what happens to the
wave packet
once it gets to large values of the $X$ coordinate.

It is also illuminating to calculate the wave function of the
 ground state  in terms of the
$X$ description. As $t\rightarrow -\infty$, the system becomes a conventional Harmonic oscillator
with constant angular frequency $\omega_0$.
Consider the   state  specified by the condition,
\be
\label{gndstate}
\hat{a}|0>=0,
\ee
which we will refer to as the ground state.
We are interested in asking how expectation values in this  state evolve with time.
We have been working in the Heisenberg picture above. It is useful to answer
this question by constructing the  wave function
 in the Schrodinger picture. The resulting time dependence of the
 wave function carries information about the time dependent expectation values for  all operators
 in this state.

As discussed in appendix \ref{sec:shm} the Schrodinger picture wave function for the ground state  is given by,
\be
\label{gndspsi}
\psi(x,t)={A \over \sqrt{f^*(t)}}e^{i[({\dot{f}\over f})^*{x^2\over 2}]},
\ee
where $A$ is a time independent constant which is fixed by requiring that the state has unit norm.
Two features of the resulting behavior of this wave function near $t=0$ are worth commenting upon.

First,  as discussed in Appendix \ref{sec:shm}, it follows that
the probability density $|\psi(x,t)|^2$ is given by,
\be
\label{probden}
|\psi(x,t)|^2={|A|^2 \over |f|}e^{-[{\omega_0 x^2 \over |f|^2}]}.
\ee
It follows from eq.(\ref{deff}), eq.(\ref{defnu}), and the
properties of Hankel functions that
\be
\label{bform}
|f|^2 \sim (-t)^{-p} \rightarrow \infty, t \rightarrow 0^-.
\ee
Thus the  probability distribution in $X$ become infinity spread out, as
$t\rightarrow 0$.
We saw above that for generic coherent states the center of the wave packet runs off to infinite $X$.
The vacuum is a non-generic coherent state, for which this does not happen. The expectation value of
$<\hat{X}>$ vanishes in this state, this corresponds to $A,B$ in eq.(\ref{csol}) both vanishing.
However we see now that even for   this state the spreading of the wave function, which
is a quantum effect,
makes the wave function sensitive to large $X$.

Second, the exponential factor in eq.(\ref{gndspsi}) gives rise to a phase factor,
\be
\label{bhnsing}
e^{i[({\dot{f} \over f})^*{x^2\over 2}]}\sim e^{[i{px^2\over 4 t}]}.
\ee
This phase factor oscillates ``wildly'' near $t=0$.
As a result $\psi(x,t)$ does not have a well defined limit as $t\rightarrow 0^{-}$.
This feature will be crucial in the subsequent analysis that follows.

We have  neglected the quartic interaction term of  eq.(\ref{lagnew}) in the analysis above.
We now see from eq.(\ref{bX})
 that
$X$ diverges, as $t^{-p/2}$, and thus $e^\Phi X^4$ goes like, $t^p t^{-2p}
 \rightarrow \infty$, as $t\rightarrow 0$, and  also diverges. This means that despite the
vanishing dilaton  the
quartic term  cannot be regarded as a small perturbation.

It is worth recalling now that the theory we started with was formulated in terms of the
$\tilde{X}$ variable, eq.(\ref{toy}). The change of variables
from $X$ to $\tilde{X}$ is in fact singular
at $t=0$ where the dilaton vanishes. Moreover, as we will see below, $\tilde{X}$ is in fact finite,
as $t\rightarrow 0$, since the
 the rate at which $X$ diverges in  eq.(\ref{nvarx}),
 is exactly the rate at which $e^{-\Phi/2}$ also blows up.
It is therefore worth constructing a description directly in terms of the $\tilde{X}$ variable.
The resulting behavior at finite $\tilde{X}$ will also  provide information about what happens at
infinite $X$, which is the additional data we seek.
We turn to this next.

\subsection{The ${\tilde{X}}$ Description}

As was mentioned above, to gain some understanding in the $\tilde{X}$ variable
 we first begin by neglecting the quartic term. The action in terms of this variable then takes the form,
\be
\label{lagtilde}
S=\int dt~ e^{-\Phi} {1\over 2}[ {\dot{\tilde X}}^2-\omega_0^2 \tilde{X}^2].
\ee
Classical solutions take the form,
\be
\label{bxtilde}
\tilde{X}(t)=e^{{\Phi \over 2}} X(t)=e^{{\Phi \over 2}} \sqrt{(-t)}[ A J_\nu(-t) + B N_\nu(-t)].
\ee
It is easy to check from eq.(\ref{defnu}) and
eq.(\ref{gendil}) that
 as $t\rightarrow 0$,
\be
\tilde{X}(t)\sim B e^{{\Phi \over 2}} N_\nu(-t) \rightarrow constant.
\ee
Thus classical trajectories do not reach  $|\tilde{X}| \rightarrow \infty$, but instead
are at  finite values as $t\rightarrow 0$.

From the relation, eq.(\ref{nvarx}), and eq.(\ref{hatx}),
it follows that the operator $\hat{\tilde{X}}$ in the
quantum theory is
\be
\label{hattildex}
\hat{\tilde{X}}={e^{\Phi/2}\over \sqrt{2\omega_0}}[\hat{a} f(t) + \hat{a}^\dagger f^*(t)].
\ee
The ground state satisfies the condition given in\footnote{Note that for the dilaton dependence given in
eq.(\ref{dil1sol2}), the Lagrangian in terms of the $\hat{\tilde{X}}$ variable  does not reduce to
that of a standard harmonic oscillator as $t\rightarrow -\infty$.
This is in contrast to the  solution, eq.(\ref{dil1sol1}), where we do
 get a standard harmonic oscillator, as $t\rightarrow -\infty$, as  discussed in appendix \ref{sec:milne}.
 It is this latter case that is better defined in any case, as was discussed
in section 3. The  ground state   in this latter case, which becomes the vacuum of the standard
harmonic oscillator in the far past,  behaves  similarly to the vacuum state consider here near the singularity.}
eq.(\ref{gndspsi}).
The resulting wave function in the Schrodinger picture (see appendix \ref{sec:shm}) is,
\be
\label{gndstilde}
\psi(\tilde{x},t)={A \over \sqrt{f^*(t) e^{\Phi\over 2} } }
e^{i\{ [({\dot{f}\over f})^* +{\dot{\Phi} \over 2}]{e^{-\Phi} \tilde{x}^2 \over 2} \}}.
\ee
The probability to find the system between ${\tilde x}, {\tilde x}+d{\tilde x}$, is
$|\psi(\tilde{x},t)|^2$, and is  given by,
\be
\label{probt}
|\psi(\tilde{x},t)|^2={|A|^2 \over \sqrt{|f|^2 e^{\Phi}}} e^{-[{\omega_0 \tilde{x}^2 \over |f|^2 e^\Phi}]}.
\ee
 From eq.(\ref{bform}), eq.(\ref{gendil}), it follows that $|f|^2 e^{\Phi}$ goes to a constant as
 $t\rightarrow 0$.  Thus $|\psi(\tilde{x},t)|^2$ becomes a well-defined smooth Gaussian function in the
 limit $t\rightarrow 0$. The absolute value of the wave function,
$|\psi(\tilde{x},t)|$ thus has a smooth limit as $t\rightarrow 0^-$.
Contrast this with the phase of the wave function.
As discussed in appendix \ref{sec:shm},
\be
\label{phase}
e^{-\Phi}[({\dot{f}\over f})^* +{\dot{\Phi} \over 2}]  \rightarrow 1/(-t)^{p-1},
\ee
as $t\rightarrow 0^-$,
and thus the phase of the wave function goes like,
\be
\label{wfphase}
e^{i[({\dot{f}\over f})^* +{\dot{\Phi} \over 2}]{e^{-\Phi} \tilde{x}^2 \over 2}} \rightarrow
e^{i[{ C {\tilde x}^2 \over (-t)^{p-1}}]},
\ee
where $C$ is a constant.
Note that for $p\ge 1$ the phase factor diverges \footnote{For $p=1$, the divergence goes
like $\log(-t)$. This follows from standard properties of Bessel  functions, and
also from the general discussion in the next subsection.}.
The result is that the wave function, eq.(\ref{gndstilde}), does not have a well defined limit as
$t\rightarrow 0^{-}$.
In terms of expectation values, this divergence result in the expectation value
for $\hat{\tilde{P}}^2$ blowing up. One finds that,
\be
\label{phat}
<\hat{\tilde{P}}^2> \sim (-t)^{2 (1-p)} \rightarrow \infty.
\ee

We have considered the ground state wave function above. In the subsection that follows,
we will give a general argument for why  the same diverging phase factor
arises for the wave function
of any state. This  general argument  will also include quartic terms.

\subsection{General Analysis of Wave Function Near $t=0$}

The  behavior  of  wave function can be analyzed quite generally in the vicinity of $t=0$.
It is easy enough to carry out the analysis in the the full  quantum mechanics system, eq.(\ref{qmsys}),
 including the
 quartic interaction terms.

The Schrodinger equation takes the form,
\be
\label{set2}
-{e^\Phi \over 2} \partial_{\tilde{x}}^2\psi+e^{-\Phi}V(\tilde{x})\psi=i\partial_t\psi,
\ee
where the potential, $V(\tilde{x})$, is
\be
\label{defvh}
V(\tilde{x})={1\over 2} \omega_0^2 \tilde{x}^2+ \tilde{x}^4.
\ee
Since $e^\Phi$ vanishes near the singularity let us begin by assuming that the potential energy term
on the lhs of eq.(\ref{set2}) dominates, we will verify below
that this assumption is self-consistently true.  This gives,
\be
\label{set3}
e^{-\Phi}V(\tilde{x})\psi \simeq i\partial_t\psi,
\ee
which can be easily solved to give,
\be
\label{sol}
\psi(x,t)=e^{-iG(t) V(\tilde{x})} \psi_0(x),
\ee
where,
\be
\label{valg}
G(t)=\int dt e^{-\Phi} =-{(-t)^{1-p} \over (1-p)},
\ee
and $\psi_0(x)$ is a time independent integration constant.
Here we have used eq.(\ref{gendil}) for the dilaton. For $p>1$, we see that $G(t)$ diverges at the
singularity, leading to a diverging  phase factor in the wave function. This divergence is
a general feature, independent of the initial state, $\psi_0(x)$.In the quadratic case where
\be
\label{quadcase}
V(\tilde{x})={1\over 2} \omega_0^2\tilde{x}^2,
\ee
we see that this phase factor agrees with what
was obtained in the exact solution for the ground state, eq.(\ref{wfphase}).

To check the self consistency of our assumptions let us evaluate the contribution due to the kinetic
energy term in eq.(\ref{set2}) on the solution, eq.(\ref{sol}). It is useful to analyze the two cases
$p>1$, and $p<1$ separately \footnote{A similar analysis can also be carried out when $p=1$.},
in both cases we see below that the kinetic energy term is subdominant compared to the potential energy
term.

For $p>1$  the leading contribution to the kinetic energy comes when the spatial derivatives act on the phase factor, and not on $\psi_0(x)$. This gives,
\be
\label{ft}
-{e^\Phi \over 2} \partial_{\tilde{x}}^2\psi  \simeq  {e^\Phi\over 2}[ G(t)^2 V'(\tilde{x})^2
+i G(t) V^{''}] \psi.
\ee
Since $G(t)$ diverges the dominant contribution comes from the first term on the rhs leading to
\be
\label{fttwo}
-{e^\Phi \over 2} \partial_{\tilde{x}}^2\psi
 \sim  {e^\Phi\over 2}[ G(t)^2 V'(\tilde{x})^2] \psi \sim
t^2 e^{-\Phi} V'(\tilde{x})^2 \psi,
\ee
where we have used the behavior for $G(t)$ in eq.(\ref{valg}).
Comparison with the potential energy term in eq.(\ref{set2}) shows that this contribution is suppressed
by an extra power of $t^2$.

For $p<1$ the leading contribution comes when the spatial derivatives act on $\psi_0(x)$. This gives,
\be
\label{plft}
-{e^\Phi \over 2} \partial_{\tilde{x}}^2\psi\simeq  -{e^\Phi \over 2} e^{-iG(t) V(\tilde{x})}
\psi_0(x)^{''}.
\ee
Comparing with the potential energy term in eq.(\ref{set2}) we see that this term is suppressed by an
 extra power of $t^{2p}$.

From the point of view of the bulk  dual cosmology, we are especially interested
in the question of whether the  state can be continued past $t=0$, with the dilaton
varying for example like,
\be
\label{extdil}
e^\Phi=|t|^p.
\ee
We have found above
that  the wave function for a general state in the quantum mechanics model does not have a well defined limit as
$t\rightarrow 0^{-}$, when $p>1$. This means  it is not meaningful  to ask about
its continuation for $t>0$ in this case.
To obtain this continuation one would need to impose that the wave function at $t=0$ is continuous,
i.e., meets the condition,
\be
\label{contcond}
\psi(\tilde{x},t=0^-)=\psi(\tilde{x},t=0^+).
\ee
This condition cannot be imposed if $Lim_{t->0^-}\psi(\tilde{x},t)$ does not exist.

\subsection{The Energy Blows up at $t\rightarrow 0$} \label{sec:enblow}
We continue with our general analysis of the wave function in the quantum mechanics system in this subsection and find that
for a generic state, and all values of $p>0$, the energy at $t\rightarrow 0^-$ diverges.
We will work below  with a general potential $V(x)$.

The Hamiltonian operator, $H$,  is given by the left hand side of the Schrodinger
equation, eq.(\ref{set2}), eq.(\ref{fse}).
We have argued above that the kinetic energy contribution is subdominant to the
potential energy, near $t=0$, so that,
\be
\label{exh}
<H>\simeq e^{-\Phi}<V>.
\ee
The expectation value of the potential,
\be
\label{exv}
<V>=\int d\tilde{x} V(\tilde{x}) \psi^*(\tilde{x},t)\psi(\tilde{x},t).
\ee
Substituting for the wave function from eq.(\ref{sol}) we see that the phase factor drops
out so that $<V>$  near $t=0$ is given by
\be
\label{exva}
<V>=\int d\tilde{x}V(\tilde{x}) |\psi_0(\tilde{x})|^2 ,
\ee
and is time independent.
This means the leading time dependence in $<H>$ comes from the prefactor $e^{-\Phi}$ in front in
eq.(\ref{exh}), leading to the conclusion that the energy diverges as
\be
\label{exph}
<H>\rightarrow (-t)^{-p}
\ee
when  $t\rightarrow 0$.
Note that this conclusion holds for all $p>0$.

This conclusion can be avoided if the state is such that  $<V>$ vanishes.
This issue
is examined further in some detail in Appendix~\ref{sec:subenergy}. The conclusion, after analyzing subleading
corrections which could also have been potentially divergent contributions, is the following:
Unless $p>2$, in which case a divergent contribution to the energy arises from the kinetic energy term,
the subdominant contributions to the energy  do not diverge as $t\rightarrow 0$.
Thus, requiring that  $<V>$ vanishes  is enough to ensure that the energy stays finite.

Now even if $<V>$ vanishes we should note that  the expectation value of $<H^2>$ will diverge.
From the discussion above it follows that
\be
\label{exh2}
<H^2>\simeq e^{-2\Phi} <V^2> .
\ee
In general in a state where  $<V>$ vanishes, $<V^2>$ will not vanish.
This means that even in those special states where the expectation value of the energy stays finite
as $t\rightarrow 0$, the fluctuations about this finite  mean value  will diverge \footnote{
 In the next section we will apply the   discussion of this and the previous subsection
 to the ${\cal N}=4$ gauge theory. In that case, both $<H>, <H^2>$
should scale like
$N^2$ - the number of colors. This means that the fluctuations in energy will be
suppressed in the large $N$ limit.  This suggests that  to leading order in $1/N$, and for $p<2$,
the vanishing of $<V>$ is sufficient to ensure that the expectation value of energy stays
finite when $t\rightarrow 0$.
}

\section{Analysis in Field Theory}\label{sec:gt}
In the previous  section we have analyzed the quantum mechanical model, eq.(\ref{qmsys}) extensively. Here we return to field theory, first discussing the toy model
field theory, eq.(\ref{toy}), and then turning to the deformed ${\cal N}=4$ field theory eq.(\ref{lagn4}).

 The lessons from the study of the quantum mechanics model can be directly applied
 to the field theory, eq.(\ref{toy}).
Carrying out the field redefinition in the full field theory gives rise to a Lagrangian, upto a surface term, 
\be
\label{lagnewre}
L= -{1\over 2}(\partial X)^2 - m^2(t) X^2 - e^{\Phi} X^4.
\ee
with $m^2$ being a tachyonic time dependent mass,
\be
\label{masstwore}
m^2= - {\alpha^2\over t^2}
\ee
with,
\be
\label{valalpha1re}
\alpha^2={p\over4}(p+2),
\ee
which diverges as $t\rightarrow 0$.

The first lesson which carries over from quantum mechanics to field theory is that contrary to  what one might have guessed at first,  the quartic term cannot be neglected
near $t=0$. The second lesson is that  the variable $X$ is not so convenient to work with and
the analysis is more conveniently carried out in terms of the original variables
$\tilde{X}$. The third lesson is that the analysis is conveniently carried out in terms of the Schrodinger picture.
This last lesson is not easy to apply in field theory, since typically the Schrodinger picture has not been used in this context.
Nevertheless with the experience of the quantum mechanics model in mind we will
in this subsection analyse the field theory in the Schrodinger picture;
various caveats will be discussed in the next subsection.

In the Schrodinger picture in field theory the state of the system is described by a time dependent wave functional,
$\psi[\tilde{X}(x),t]$, which  satisfies the Schrodinger equation,
\be
\label{fse}
-{1\over 2} \int d^3x~
e^\Phi {\delta^2 \psi \over \delta{\tilde{X}}^2} + e^{-\Phi}
 V[\tilde{X}] \psi = i\partial_t \psi.
\ee
Here the potential energy, $V[\tilde{X}]$, is a functional given by,
\be
\label{potfunc}
V[\tilde{X}]=\int d^3 x\{ {1\over 2} (\partial_i \tilde{X})^2 +\tilde{X}^4 \}.
\ee

The first term on the l.h.s. in eq.(\ref{fse}) is the kinetic energy, and the second term is the potential energy. 
We see that, like in quantum mechanics, the kinetic energy term has a prefactor
$e^\Phi$, while the potential energy has the prefactor, $e^{-\Phi}$.
This suggests that the
kinetic energy term is once again subdominant close to the singularity, leading to the solution,
\be
\label{wfunctional}
\psi[\tilde{X}(x^i),t]=e^{-i\{G(t) V[\tilde{X}]\}}\psi_0[\tilde{X}(x^i)].
\ee
We see that the wave functional has a phase factor which diverges, as in the quantum mechanics case,  resulting in a singular limit for the wave function if $p\ge 1$.
Moreover with the kinetic energy being subdominant, near $t=0$, the Hamiltonian is well approximanted by,
\be
\label{haft}
<H>\simeq e^{-\Phi}<V>.
\ee
As is the quantum mechanics model the phase factor drops out in the expectation value of $<V>$ near $t=0$ leading to the conclusion that in field theory as well,
$<H>$ goes like, eq.(\ref{exph}), and therefore blows up for all $p>0$.

The central assumption here is that the kinetic energy is subdominant compared to
the potential energy near $t=0$. This was shown to be self consistently true in the case of quantum mechanics. We will analyse this issue for the field theory in the next subsection.

Before proceeding let us also mention that  we analyze the quadratic theory in further
 detail in appendix~\ref{sec:partprod}.
We consider a dilaton profile of the form, $e^\Phi=|t|^p$, and evolve the field theory
in this background, starting from the vacuum into the far future. It is useful for this purpose
to regulate the dilaton profile near $t=0$ in a manner we make more precise in the appendix.
Our conclusion is that particle production always occurs. For $p<1$ the particle production is finite,
while for $p>1$ it becomes infinite as the regulator is taken to zero.

Turning now to the gauge theory, eq.(\ref{lagn4}), we see that
the coupling of the dilaton to the quartic scalar potential and the fermionic Yukawa terms
are proportional to positive powers of $e^\Phi$ and can be neglected when the dilaton is very small.
The Fermions and scalars have canonical kinetic terms.
The gauge field in contrast has a non-canonical kinetic energy term, it is the analogue
of the $\tilde{X}$ field in the toy model, eq.(\ref{toy}).
As $t\rightarrow 0$, and the dilaton vanishes, it is this gauge kinetic energy term
which will determine the behavior of the system.
Accordingly, in the analysis below we focus on the pure gauge theory, without fermions and scalars, and
with action,
\be
\label{gflag}
S=\int d^4x (-{1\over 4 e^{\Phi}}) \Tr \  F_{\mu\nu}F^{\mu\nu}.
\ee

The equation of motion is,
\be
\label{egf}
D_\mu(e^{-\Phi} F^{\mu\nu})=0.
\ee
We work in Coulomb gauge, where,
\be
\label{cgauge}
A_0=0.
\ee

Consider now the non-interacting theory.
The equation of motion, eq.(\ref{egf}), for, $\nu=0$, becomes, (this is the Gauss Law constraint),
\be
\label{glawc}
\partial_0(\partial^j A^j)=0.
\ee
Thus the longitudinal part of the gauge field is time independent.
We can now do an additional time independent gauge transformation to set
\be
\label{long}
\partial_jA^j=0.
\ee
The equation of motion, eq.(\ref{egf}), with $\nu=i$, then become,
\be
\label{etra}
\partial_0(e^{-\Phi}\partial^0 A^i)+e^{-\Phi}\partial_j\partial^jA^i=0,
\ee
for the two transverse components satisfying, eq.(\ref{long}).
Eq.(\ref{etra}) is exactly the equation we have for a scalar field with Lagrangian,
\be
\label{lagsc}
L=e^{-\Phi}(\partial \tilde{X})^2.
\ee
Thus at the quadratic level, the analysis for the gauge field reduces to that of the scalar field
considered above. There are two transverse components coming from each gauge field. We have neglected
the color degrees of freedom above. They are easily incorporated and give  $(N^2-1)$  degrees
of freedom for each of the two transverse components.

Interactions give rise to cubic and quartic terms. In Coulomb gauge, these terms do
not depend on any time derivatives. Thus, in the Hamiltonian they only contribute to the Potential energy
and not to the Kinetic energy. The interactions
 can therefore be included in a way very similar to the quartic terms
in  the toy model for $\tilde{X}$.

 In particular, we are interested in the wave function in the
Schrodinger picture. In this picture the operators are time independent.
The total potential energy is given by energy due to the magnetic field,
\be
\label{potga}
V[A_i(x)]=\int d^3x {1\over 4}Tr(F_{ij}F^{ij}).
\ee

Motivated by the quantum mechanics model on the previous section, and as in the
scalar field theory above, we now take the potential energy to dominate in the
Schrodinger equation near the singularity.
As a result the wave function
has a phase given by,
\be
\label{pgauge}
\psi[A_i(x),t]=e^{-i\{G(t) V[A_i]\}} \psi_0[A_i(x)].
\ee
The phase factor is identical to  that found in eq.(\ref{sol}) and
 diverges, as $t\rightarrow 0$,  if $p>1$,
 resulting in the wave function being singular at $t\rightarrow 0$.
The wave function above can be regarded as being dependent on only the transverse components of the
vector potential. Alternatively, we can take  the wave function to be dependent on a general
 gauge potential (with $A_0=0$), and then impose Gauss' law,
\be
\label{glaw2}
\partial_i {\delta \psi\over \delta A_i}=0,
\ee
on it.

Similarly one can calculate the expectation value of the energy near $t=0$.
It has the  same form as in eq.(\ref{exph}), with the expectation value of the
potential energy now being given by,
\be
\label{poten}
<V>=\int DA_i |\psi_0[A_i(x)]|^2 V[A_i(x)],
\ee
where $V[A_i(x)]$ is given in eq.(\ref{potga}).
We see that the energy diverges  in the gauge theory as well, as $(-t)^{(-p)}$, for all $p$,
 unless the state is such that $<V>$ vanishes.

We conclude this section with two important comments.
First, naively one would have thought that as the dilaton becomes small perturbation theory should become
a good approximation. However we have seen in our analysis of the toy model  in the last section that this
is  in fact not true. In the $X$ description for the toy model,
the time dependent dilaton drives the system
to large values of $X$ resulting in the quartic term being non-negligible.
A similar argument also holds in the gauge theory. The cubic and quartic interactions  terms
are not small near $t=0$,
and as a result perturbation theory is not a good approximation.
Secondly, it should be emphasized that our analysis is valid for
finite $N$ and finite $g_{YM}^2N$. In fact from eq.(\ref{pgauge}) we see that the behavior of the wavefunctional near $t = 0$ is essentially determined by $G(t)$ and is independent of the details of the potential $V[A_i(x)]$. In the dual closed string theory, this means that this conclusion is valid in the presence of string and quantum corrections.

\subsection{A more critical look at the Field Theory Analysis}
The central assumption in the field theory discussion above  was that in the
 Schrodinger equation the potential energy which scales like $e^{-\Phi}$ dominates over the kinetic energy,
which has a prefactor $e^\Phi$ in front of it.
This assumption was motivated by our earlier analysis in quantum mechanics.
However, field theory differs from quantum mechanics in having an infinite number of degrees of freedom,  
and one might worry that this introduces additional subtleties and complications 
\footnote{We are grateful to David Gross in particular for  emphasising this point to us.}.
We turn to an examination of these issues below.

The wave function for the ground state of the Harmonic oscillator with action
eq.(\ref{lagsmt})  in the $\tilde{x}$ description is given by eq.(\ref{gndstilde}).
This is an exact result. If the potential energy is dominant the wave function
has the form, eq.(\ref{sol}). From the exact result we can ask how close to
 $t=0$ must one come for this  approximation to become a good one.
 The phase factor in the wave function eq.(\ref{gndstilde})
 contains the factor, $({\dot{f}\over f})^*+{\dot{\Phi}\over 2}$. As discussed in eq.(\ref{valp})  
of Appendix B this phase factor has a power series expansion in $t\omega_0$, near $t=0$, of form,
 \be
 \label{expa}
 ({\dot{f}\over f})^*+{\dot{\Phi}\over 2} ={1\over t}({1 -2 \nu+p \over 2}) + 2 c_2 \omega_0^2 t+ \cdots.
 \ee
 The ellipses stand for terms which are supressed when
 \be
 \label{approx}
 t\omega_0\ll 1.
 \ee
 Now the first term in eq.(\ref{expa}) vanishes due to eq.(\ref{defnu}). The
 second term is therefore the leading one and it is easy to see that this gives agreement with eq.(\ref{sol}).

 The conclusion is that the potential energy dominates over the kinetic energy term when $t$ is small enough and satisfies the condition in  eq.(\ref{approx}).
 We can also understand this on the basis of the general arguments given in section 3 for the self consistency of this approximation.
 From eq.(\ref{fttwo}) we see that the KE term goes like,
 \be
\label{fttwoa}
-{e^\Phi \over 2} \partial_{\tilde{x}}^2\psi
 \sim
t^2 e^{-\Phi} V'(\tilde{x})^2 \psi\sim t^2 e^{-\Phi}(\omega_0^2 \tilde{x})^2 \psi,
\ee
 while the potential energy term is,
 \be
 \label{pote}
 e^{-\Phi}V(\tilde{x}) \psi \sim e^{-\Phi}(\omega_0\tilde{x})^2 \psi.
 \ee
 Thus for the latter to dominate, eq.(\ref{approx}) must be true.

 Consider now a field theory in the quadratic approximation, since there are an infinite number of modes, for any non-zero and arbitrarily small time $t$,  there will always
 be some modes with high enough frequency for which the condition eq.(\ref{approx})
 is not met and thus for which the wave functional will not be well approximated by eq.(\ref{wfunctional}), or eq.(\ref{pgauge}). Including interactions will couple these modes to  low-momentum frequency
 potentially making the wave functional for all modes to be different from eq.(\ref{wfunctional}), eq.(\ref{pgauge}).

 It is of course well known that  in dealing with the infinite numbers of degrees of freedom in a field theory it is first useful to introduce a cut-off or regulator, which makes the number of degrees of freedom finite and then ask what happens as the cut-off is removed. For ease of discussion consider a momentum space cut-off $\Lambda$ ( in the gauge theory one needs a more sophisticated regulator to preserve gauge invariance, but this will
 not change the essential points in our discussion). The above analysis
 suggests  that if we take the time $t$ to vanish while keeping $\Lambda$ fixed and finite, then the potential energy should  dominate for times
 $t$ meeting the condition
 \be
 \label{deftsmall}
 t\Lambda \ll 1
 \ee
 and our conclusions in the previous section will be correct near $t=0$.
 More generally one might expect that these conclusions are  valid as long as we take $t\rightarrow 0$, before we take, $\Lambda \rightarrow \infty$.

 Unfortunately, it is not easy to make this argument precise.
 Additional complications can arise in field theory due to the effects of renormalisation \footnote{We thank the referee for emphasising this point to us.}. Usually renormalisation in field theory is discussed in terms of an effective Lagrangian which changes under RG flow. For our purposes in the discussion above the Schrodinger picture has been more useful.
 Including the effects of renormalisation   in the Schrodinger picture though  is a complicated issue that we have not fully sorted out. Presumably the Hamiltonian which governs the time evolution of the wave functional needs to be made well defined by appropriate operator  ordering and this introduces the effects of renormalisation.

 One would expect that at least some of the consequences of renormalisation can be
 incorporated by first constructing an effective Lagrangian by integrating out  high frequency modes and then using this effective Lagrangian to  construct a Hamiltonian that governs
 the time evolution of the surviving low-frequency modes.
 Since the ${\cal N}=4$ theory is conformally invariant any renormalisation
 of the effective Lagrangian must be due to the time dependence of the dilaton and thus operators which are induced by this renormalisation must have coefficients proportional to derivatives of the dilaton. In turn such operators could then also
 change the Hamiltonian resulting in extra operators in it with coefficients proportional to time derivatives of the dilaton.

The essential reason why in our discussion the potential energy dominates is that it scales like
$$V\sim e^{-\Phi},$$
whereas the kinetic energy term has a prefactor $e^\Phi$ in front of it that supresses it.
However suppose as an example of the consequences of renormalisation the
potential energy acquires an extra term  which arises at one loop so that it now has the form,
$$V=e^{-\Phi}[{\cal O}_1 +e^{\Phi} {\dot{\Phi}^2 \over \Lambda^2} {\cal O}_2]. $$
Here ${\cal O}_1$ is the operator corresponding to the magnetic field energy,
$\Lambda$ is the cutoff scale  and
${\cal O}_2$ is the additional operator which arises at one loop.
If the condition
\be
\label{condpa}
p-2<0
\ee
is met, this second term could get important close to $t=0$.
This will mean that one has to include additional loop effects that arise
beyond one loop as well. Resumming these effects could well lead to
a much smaller potential energy. For example, if these corrections 
take the form 
of a geometric series, we would get schematically
$$V \sim e^{-\Phi}[1+ e^{\Phi}{ (\dot{\Phi})^2 \over \Lambda^2}+ 
({e^{\Phi}(\dot{\Phi})^2 \over \Lambda^2})^2) + \cdots ]$$
we would get after resumming
\be
\label{condare}
V \sim 
e^{-\Phi}[{\Lambda^2\over \Lambda^2-e^\Phi (\dot{\Phi})^2}]\sim {\Lambda^2 e^{-2\Phi}\over (\dot{\Phi})^2},
\ee
sufficiently close to $t=0$, if the condition eq.(\ref{condpa}) is met. We see that the effects of renormalisation can therefore supress the potential energy term. In particular if
\be
\label{compre}
e^{-2\Phi} {\Lambda^2\over \dot{\Phi}^2} \sim e^{\Phi}
\ee
this supression would make the the Kinetic energy term comparable. It is easy to see that eq.(\ref{compre}) will be met for small enough time,$t$, if,
\be
\label{condrej}
p<2/3.
\ee

The summary is that our analysis in the field theory is not complete and our conclusions about the gauge theory being singular should be taken as being 
suggestive but not  conclusive.
To analyse the gauge theory in a well-defined manner one must introduce a regulator. Once this is done a very rich set of counter terms are allowed in the process of renormalisation, and such counterterms can potentially invalidate our conclusions. This would happen if they  make the potential energy comparable to the kinetic energy or even smaller than it near $t=0$ thereby significantly changing the form of the wave function and potentially making the gauge theory non-singular.
Whether this happens or not requires a detailed understanding of renormalisation in the gauge theory in these time dependent background. This is a fairly complicated
subject and we leave it for the future.

We end with some comments. Introducing a UV regulator in the gauge theory is dual to introducing a boundary in the bulk that regulates the IR behaviour. 
Some consequences of renormalisation, in the null dependent case, have been worked out in \cite{Chu:2006pa}, where indeed terms in the Wilsonian action
with coefficients proportional to ${ (\Phi')^2\over \Lambda^2}$ were found.

\section{Cosmological Solutions} \label{sec:cosmosol}

The motivation for this investigation came from trying to understand
some cosmological solutions \cite{DMNT1,DMNT2,ADNT}.
These solutions can be thought of
as deformations of $AdS_5\times S^5$ and have a dual description in
terms of the ${\cal N}=4$ theory with a time dependent dilaton. As the
dilaton becomes small on the boundary the bulk curvature becomes
larger and larger, eventually becoming singular at $t=0$.  In this
section we ask what the above analysis in the gauge theory teaches us
about these cosmological solutions.

We begin with a brief review of these solutions and then return to the
gauge theory later.

\subsection{The Gravity  Solutions}

The solutions  arise in IIB theory  and are
 deformations of $AdS_5\times S^5$. The $S^5$ factor is
unchanged in the deformations, and accordingly we will omit it below
and only discuss the solution in the remaining five dimensions.

The first solution we consider has  the   5-dimensional metric,
\be
\label{metric1sol1}
ds^2={1\over z^2}\bigl[dz^2+|Sinh(2t)|[-dt^2+{dr^2\over 1+r^2}+r^2(d\theta^2+\sin^2\theta d\phi^2)]
\bigr],
\ee
and dilaton,
\be
\label{dil1sol1}
e^{\phi(t)}=g_s |\tanh t |^{\sqrt{3}}.
\ee
This solution was discussed in \cite{ADNT}.

In the far past, as $t\rightarrow -\infty$, the dilaton goes to a constant, and the metric becomes
$AdS_5$. One can see this by going  to coordinates,
\be
\label{coordinates}
r={R\over \sqrt{\eta^2-r^2}}, \ \ e^{-t}=\sqrt{\eta^2-R^2},
\ee
in which the metric, eq.(\ref{metric1sol1}),  and dilaton, eq.(\ref{dil1sol1}), take the form,
\be
\label{metric2sol1}
ds^2={1\over z^2}\bigl[dz^2+|1-{1\over (\eta^2-R^2)^2}|
[-d\eta^2+dR^2+R^2d\Omega_2^2] \bigr].
\ee
The far past, $t\rightarrow -\infty$, corresponds to $(\eta^2-R^2)
 \rightarrow \infty$, it is clear from eq.(\ref{metric2sol1}) that the
 metric asymptotes to $AdS_5$ in this limit.  The dilaton in these
 coordinates is,
\be
\label{dil2sol1}
e^{\phi}=|{\eta^2-R^2-1\over \eta^2-R^2+1}|^{\sqrt{3}}.
\ee
At $t=0$ the solution, eq.(\ref{metric1sol1}) has a singularity. The curvature scalar diverges
like $R \sim {1\over t^3}$ as $t\rightarrow 0$.

The dilaton vanishes as $t\rightarrow 0$, eq.(\ref{dil1sol1}). Thus the singularity occurs at weak
string coupling. This singularity is the main focus of our analysis.

The region $t<0$, which is the  region of spacetime  before the singularity,
 maps to $\eta^2-R^2 >1$, in the $(\eta, R)$ coordinates, while the singularity, which is at $t=0$,
maps to the locus, $\eta^2-R^2=1$.

Another 5-dim solution is given by
\be
\label{metric1sol2}
ds^2={1\over z^2}[dz^2+|2t|[-dt^2+(dx^1)^2 + (dx^2)^2+(dx^3)^2],
\ee
with dilaton,
\be
\label{dil1sol2}
e^{\Phi(t)}=g_s |t|^{\sqrt{3}}.
\ee
This solution does not asymptote to $AdS_5$ in the far past, as
$t\rightarrow -\infty$.  However its behavior at the singularity, as
$t\rightarrow 0$, is very similar to the solution discussed above,
eq.(\ref{metric1sol1}), eq.(\ref{dil1sol1}).  The $z=const$
hypersurfaces in both metrics are of the FRW form. The difference
between the two metrics is that this 4 dim.  FRW cosmology has
constant negative curvature in the first case, eq.(\ref{metric1sol1})
while it is flat in the second case, eq.(\ref{metric1sol2}). This
difference is increasingly unimportant near the singularity, where the
dominant source of stress energy is provided by the diverging time
derivative of the dilaton, rather than the spatial curvature.  Since
the dilaton is essentially identical near the singularity, at
$t\rightarrow 0$, in both cases, the resulting spacetimes also are
essentially the same.

In section~\ref{sec:bklcosmo} we will explore some additional cosmological
  solutions and comment on their gauge theory duals.
Some of these solutions differ from the two solutions discussed above at early times but their
behavior near $t=0$ becomes the same as in the solutions above.

\subsection{The Gauge Theory Duals}
For purposes of studying the Field theory dual, we start  with the  first
bulk solution considered above, eq.(\ref{metric1sol1}), eq.(\ref{dil1sol1}), or equivalently
eq.(\ref{metric2sol1}), eq.(\ref{dil2sol1}). This solution asymptotes
  in the far past  to $AdS_5$ with a constant dilaton. As discussed in
\cite{DMNT1}, \cite{DMNT2}, this corresponds to starting in the far past with the $N=4$
SYM theory in the vacuum state.
The spacetime, eq.(\ref{metric1sol1}), has a boundary at $z\rightarrow 0$. We see
 that  the metric on the boundary is
conformal to flat space. As was discussed in \cite{ADNT},  we can then take the metric of the spacetime
in which the dual gauge theory lives to be the  $4$ dim. Minkowski-space metric.

Since this is an important point
 let us pause to briefly comment on it further.
In general  a Weyl transformation in the boundary theory corresponds
to a Penrose-Brown-Henneaux (PBH) transformation  in the bulk.  The explicit PBH transformation
which gives rise to a flat boundary metric for eq.(\ref{metric1sol1})
 was found in \cite{ADNT}.
The metric in eq.(\ref{metric1sol1}) has a second order pole as  $z\rightarrow 0$.
The  metric on the boundary defined by
\be
\label{bmetric}
ds_4^2\equiv Lim_{z \rightarrow 0} z^2 g_{\mu\nu}dx^\mu dx^\nu,
\ee
where $x^\mu$ denotes coordinates on the boundary, is,
\be
\label{bmet2}
ds_4^2= |Sinh(t)| [-dt^2+{dr^2\over 1+r^2} + r^2 (d\theta^2 + \sin^2\theta d\phi^2)].
\ee
After the PBH transformation the resulting 4 dimensional metric
is  given by,
\be
\label{bmet3}
ds_4^2=e^{-2t}[-dt^2+{dr^2\over 1+r^2} + r^2 (d\theta^2 + \sin^2\theta d\phi^2)].
\ee
This is in fact flat space in Milne coordinates. It is easy to see this. The coordinate transformation,
eq.(\ref{coordinates}), turns this metric into the familiar Minkowski metric,
\be
\label{bmet4}
ds_4^2=-d\eta^2+dR^2+R^2d\Omega^2
\ee
The region to the past of  the singularity, is given by, $-\infty<t<0$,
 in eq. (\ref{bmet3}),
and  maps  to the region
$\eta^2-R^2>1$, which is part of one of the  Milne wedges. The rest of this Milne wedge is given by
$0<t<\infty$, to which the metric eq.(\ref{bmet3}) automatically extends. The
boundary of the Milne wedge lies at $t\rightarrow \infty$. Starting from $t\rightarrow -\infty$
one arrives at the   singularity,  at $t=0$, before reaching the boundary of the Milne wedge.

The dual gauge theory knows about the time dependence of the bulk through the  varying dilaton.
The exponential of the  dilaton is equal to the coupling constant of the Yang Mills theory, eq.(\ref{dil}).
Thus the varying dilaton gives rise to a varying Yang Mills coupling.
The dilaton depends on the Milne time coordinate, $t$, eq.(\ref{bmet3}), and takes the form,
eq.(\ref{dil1sol1}).

In summary we see that the dual gauge theory to this cosmological solution lives in flat space
with a varying dilaton which vanishes at the singularity. The analysis in the preceding sections
can be now be used to determine the nature of this singularity.
We turn to this in the next subsection below.

Before that, it is also useful to discuss the dual to the second cosmological solution introduced
in the previous
subsection, with metric and dilaton given by, eq.(\ref{metric1sol2}), eq.(\ref{dil1sol2}). In this
case the dual gauge theory also lives in flat space, but the dilaton depends on Minkowski time instead of Milne time. To see this note that
from eq.(\ref{metric1sol2}) it follows that  the  boundary metric as defined by eq.(\ref{bmetric})
is conformally flat.
After a suitable  PBH transformation the boundary metric  becomes
 that of flat Minkowski space,
\be
\label{bmetsol2}
ds^2=-dt^2+(dx^1)^2 + (dx^2)^2+(dx^3)^2.
\eeq
The dilaton dependence is given in eq.(\ref{dil1sol2}), we see that it is only a function
of Minkowski time.

The first solution, eq.(\ref{dil1sol1}),
has a dilaton which goes to a constant in the far past,
and the dual gauge theory starts in the vacuum state of the $N=4$ theory as $t\rightarrow -\infty$.
In contrast in the second solution, eq.(\ref{dil1sol2}),
 the dilaton blows up in the far past, this makes the dual
map to the boundary theory less clear.
It will turn out that the behavior at the singularity of the two theories is similar \footnote{
This similarity is in parallel with the fact, mentioned in the previous section,
that the two corresponding  bulk solutions also behave similarly near the singularity.},
near the singularity,
and a little easier to analyze in the second case, where the variation is with respect to Minkowski time.
For this reason, and for the limited purpose of asking questions near the singularity,  we will focus
some of the following discussion on the second hologram.

\subsection{Gauge Theory and Gravity}
In this section we relate what was learnt in the analysis of the gauge theory above to the specific
cosmological solutions of interest.

We see from eq.(\ref{dil1sol1}), eq.(\ref{dil1sol2}),  that the solutions correspond to
\be
\label{valpgrav}
p=\sqrt{3}
\ee
In particular this means that for these solutions, $p>1$.
We now see that, with the provisos discussed above,
the analysis in the gauge theory suggests that
 the system  becomes  genuinely sick as  $t\rightarrow 0^-$, . In particular the wave function in the Schrodinger
picture, acquires a wildly oscillating phase, eq.(\ref{sol}), and thus does not have
a good limit, as $t\rightarrow 0$. This also means the state cannot be sensibly continued past $t=0$. We remind the reader that these conclusions are not definitive, in particular  various caveats discussed in section 4.1 apply here,
and there are additional issues, having to do with renormalisation that we have not adequately discussed in this work.

If true, the conclusions should hold
 regardless of the state of the system.
In the far past, for the solution, eq.(\ref{metric1sol1}), eq.(\ref{dil1sol1}), the state is known to be the
vacuum of the ${\cal N}=4$ theory. However,  as time progresses and the dilaton becomes smaller the state evolves.
The analysis in the gauge theory leading to the wave function of form, eq.(\ref{pgauge}),
 is only valid very close
to $t=0$, where the state would be different in general from the  vacuum.

It also follows from the gauge theory, eq.(\ref{exph}),
 that   the energy diverges like,
$(-t)^{-\sqrt{3}}$, as
$t\rightarrow 0$. If our approximations hold, the only way to avoid this conclusion would be if 
  the system, which starts in the vacuum state in the far past, evolves to a  non-generic state
 near $t=0$ for which $<V>$, eq.(\ref{exv}), vanishes.
 While this seems unlikely, since such states are non-generic, one cannot rule
out this possibility. 
 Note however that even in this case the discussion of the previous
two paragraphs would continue to hold and the system would become  sick as $t\rightarrow 0^-$.
Also, as was mentioned in section~\ref{sec:da}, even in such a non-generic state the fluctuations in the energy,
which are suppressed at large $N$ would still diverge and  the gauge theory at finite $N$ would still be singular.

The cosmological solution in eq.(\ref{metric1sol1}), eq.(\ref{dil1sol1}) maps to a boundary theory
where the dilaton depends on Milne time, rather than Minkowski time. This does not make an essential difference to
the analysis in the gauge theory which in the above sections was carried out for the dilaton
being a function of Minkowski time. This issue is analyzed further in the appendix~\ref{sec:milne}.
 The Milne case correspond to having a non-trivial metric in the boundary theory.
The dilaton vanishes at the point $t=0$ which is perfectly smooth in Milne coordinates, thus the
nontrivial metric does not make an essential difference to the discussion of the singularity.

More generally one can consider other cosmological solutions, which are different in the far
past, but which also behave like the two examples discussed above, near $t=0$ where the dilaton
vanishes.   In all these cases, as long as the metric on the boundary is well behaved at $t=0$ the
above discussion should apply.
In  section~\ref{sec:bklcosmo}, we give examples of BKL cosmologies, with the dilaton, which at late times
asymptote to the example, eq.(\ref{metric1sol2}), eq.(\ref{dil1sol2}). It follows from the discussion above that, subject to the caveats mentioned above,  the
 singularity in all these solutions  then is a genuine sickness of the theory.

\subsection{Concluding Comments}
We conclude this section with some more comments on the relation between the gravity solutions and
their  gauge theory description.

For the supergravity solutions, eq.(\ref{metric1sol1}), eq.(\ref{dil1sol1}),
eq.(\ref{metric1sol2}, eq.(\ref{dil1sol2}),
the stress energy tensor was calculated  in Section 5 of \cite{ADNT}.
For small $t$ the stress tensor diverges like,
\be
\label{stressten}
T_{\mu\nu} \sim {N^2 \over t^4}.
\ee
This calculation was made in the super gravity approximation which breaks down when the dilaton becomes small enough.
At very small values of the dilaton the gauge theory analysis carried out in section 4  becomes valid, subject to the various caveats discussed in section 4.1.
We have seen in section~\ref{sec:gt} that the energy density according to this analysis  goes like \footnote{
The initial state in the far past is the vacuum which is translationally invariant. Also, the dilaton
only depends on time and does not break this translational symmetry. Thus the potential energy in
eq.(\ref{potga})
scales like the volume, leading in turn from eq.(\ref{exh})
 to an energy density which is finite and
given by eq.(\ref{enden}.}
\be
\label{enden}
<\rho> \sim N^2/|t|^p.
\ee
(The factor of $N^2$ was not explicitly displayed  in the discussion  in
section~\ref{sec:gt}. It is easy to see that in the quadratic approximation, where the
gauge theory is free, the $N^2$ independent color degrees of freedom, in the presence of the
 time dependent dilaton,  give rise to an energy density that scales
in this way with $N$.  In the presence of interactions, the energy density should
continue to scale like $N^2$,  to leading order in $N$.).
Since $p=\sqrt{3}<4$ we see that the growth of energy  in the regime where the dilaton is very small,
is much slower than that in supergravity.

 In the cosmological solutions we are studying here,
the energy being pumped in by the dilaton does not lead to the formation of a black hole
 in the regime  where supergravity is valid, for $t<0$. This means that in the dual gauge theory the energy
being pumped in is not thermalizing. To understand this we note that
in the strongly coupled gauge theory, which is dual to supergravity, the thermalization
time scale $\tau$, is expected to be of order
\be
\label{thermat}
\tau\sim {1 \over T} \sim ({N^2\over \rho})^{1/4},
\ee
where $\rho$ is the energy density, and $T$ is the temperature.
From  stress energy tensor, eq.(\ref{stressten}), we see that\footnote{
Note that eq.(\ref{stressten}) is valid when $|t|\ll 1$ for the solution eq.(\ref{metric1sol1}).
This is consistent
with the supergravity approximation being valid, as follows from eq.(\ref{scurv}).}
\be
\label{esttau}
\tau \sim t.
\ee
Thermalization would occur if the external source is pumping in energy on a time scale much slower than $\tau$.
In the situation at hand, this time scale  is of order,
\be
\label{thermathree}
t_{pump}\sim {\rho\over \dot{\rho}} .
\ee
Using eq.(\ref{stressten}) we see that $t_{pump}$ is also of order $t$. Thus the rate at which the
dilaton is pumping in energy equals the thermalization or relaxation rate of the system. This explains
why thermalization does not happen and in the dual a black hole does not form.
Another way to see this is that from eq.(\ref{thermat}), eq.(\ref{esttau}) it follows that,
\be
\label{ratt}
{\dot{T}\over T^2}\sim O(1),
\ee
so that the temperature changes too rapidly for thermalization.

We have seen that within our approximations, a dilaton which decreases all the way to zero results 
in a singular gauge
theory. Effects of renormalization could, in principle, tame the singularity.
In any case, if the dilaton profile is modified such that $e^\Phi$ never vanishes, but can
beome small at $t=0$, a smooth time evolution beyond this time is possible.
Let us consider  a situation where the dilaton varies in a smooth fashion reaching
 a minimum value  and then increasing again,
approaching a constant  in the far future which  corresponds to a large value for the 'tHooft coupling.
The time dependent dilaton  will typically lead to the  system having some non-zero energy density
 in the far future, and one would expect that given enough time this energy would thermalize.
This suggests that in the dual closed string description supergravity will
 eventually become a good approximation in the far future, and the dual geometry in the far future
will be that of a black hole in $AdS_5$.

An interesting question to  ask is how the formation of the black hole depends on the dilaton time variation.
In particular whether the black hole is always a good description of the geometry by the time the
supergravity approximation becomes good in the future, or whether for a  suitable dilaton time profile
the formation of the black hole can be delayed, till much after the supergravity description becomes valid.
This question is difficult to answer with our current level of understanding. In particular,
in  the course of the time evolution
of the dilaton it goes through a region where the 'tHooft coupling is of order unity. This region is very difficult
to analyze, since neither the gravity nor the gauge theory descriptions are  tractable then.

Finally, we have not been very precise in the discussion above about exactly  when the supergravity
approximation  breaks down. The    metric for the solutions we are considering is  in
 eq.(\ref{metric1sol1}), eq.(\ref{metric1sol2}). For  small $t$  the curvature
goes like,
\be
\label{curv}
R_{curv}={z^2\over (-t)^3}.
\ee
In this expression we have set
\be
\label{defrads}
R_{AdS}=(4\pi g_s N)^{1/4}
\ee
to unity, and
  we are also working in $5$ dim. Einstein
frame.  After taking this into account one finds that the string frame curvature is given by,
\be
\label{scurv}
R^{string}_{curv}\sim {1 \over (-t)^{(3-{\sqrt{3} \over 2 })}} {z^2\over R_{AdS}^2}.
\ee
As $t\rightarrow 0^-$ we see that the curvature blows up, resulting in a singularity
as we have discussed above.
The supergravity approximation breaks down when the curvature gets to be of order the string scale.
We see  that when this condition becomes valid depends on the radial coordinate $z$.
In particular, for all finite $t$, $z\rightarrow \infty$ has diverging curvature. This point corresponds to the
locus where the past and future horizons of $AdS_5$ (in the Poincare coordinates we are using)
intersect. The significance of this curvature singularity is unclear to us \footnote{Note that the
past Poincare horizon at $z\rightarrow \infty$, $t\rightarrow -\infty$,  with $|z/t|$ held fixed
is non-singular, as was discussed in \cite{ADNT}.}.
As time increases, the region with high curvature
(of  order string scale or more)
grows, moving to smaller $z$,  eventually leading  to a singularity for
all $z$ as $t\rightarrow 0$.

The calculation of curvature, eq.(\ref{scurv}) makes it clear that the singularity (in string frame) is due to two effects which are tied together in these cosmological solutions.
First, the dilaton vanishes causing the string frame curvature to blow up. Second, an infinite
amount of energy is dumped into the system due to the vanishing dilaton source.
If the dilaton profile is altered so that it attains a non-vanishing minimum value at
$t=0$, the total energy put into the system would be finite. If the minimum value of the
 dilaton is small enough one expects that the curvature in string units  at $t=0$ still becomes
much bigger than unity, resulting in the breakdown of supergravity.

\section{Behavior Near Singularities and BKL Cosmologies}
\label{sec:bklcosmo}

We saw in section~\ref{sec:cosmosol} that the two solutions which
were considered had very similar behavior near the  $t=0$
singularity. This was because the stress energy near the
singularity was dominated by the dilaton which had essentially the
same behavior near $t=0$ in these two cases.

The behavior of the dilaton
near the singularity is in fact shared by a much larger
class of cosmological solutions. The key point is that the Einstein
frame metric for the class of
solutions we have considered may be written in suitable coordinates as
\cite{DMNT1}
\ben
ds^2 = \frac{1}{z^2} \left[ dz^2 + \tg_{\mu\nu}(x) dx^\mu dx^\nu \right] +
d\Omega_5^2
\label{gensol}
\een
This solves the 10 dimensional supergravity equations provided
 \bea
{{\tilde{R}}}_{\mu\nu} -
{1\over 2} {\nabla}_\mu\Phi {\nabla}_\nu\Phi&=&0 \label{efe}, \nn
\\
{1 \over \sqrt{-{\tg}}} \partial_\mu (\sqrt{-{{\tg}}}\ {{\tg}}^{\mu \nu}
\partial_\nu \Phi) &=& 0
\label{deom}.
\eea
and the five form field strength is standard
\ben
F = \omega_5 + \star \omega_5
\een
In other words, any solution of 3+1 dimensional dilaton gravity may be
lifted to a solution of 10 dimensional supergravity. This means that
we can use the well known analysis of Belinski, Lifshitz and
Khlalatnikov (BKL) and subsequent work \cite{Landau}-\cite{DH:2000}
to make useful statements about
AdS cosmologies.

For instance, we can consider a cosmological solution where the
spatial 3-metric is one of the general homogenous spaces in the
Bianchi classification (see \eg\ \cite{Landau} for a lucid treatment
in the 4-dim context), with vanishing $g_{0,\al}$ components:
\be\label{genhommet}
ds_4^2 = -dt^2 + \eta_{ab}(t) (e^a_\alpha dx^\alpha) (e^b_\beta dx^\beta)\ ,
\ee
where $(e^a_\alpha dx^\alpha)$ are a triad of 1-forms defining
symmetry directions\footnote{Starting with the 1-form triad
$(e^a_\alpha dx^\alpha)$, $a$ labelling the vectors in the triad,  we
can obtain the dual vectors $e_a^\alpha$, satisfying
$e^a.e_b=\delta^a_b$. Then the symmetry algebra acting on the
homogenous space (\ie\ the spatial metric) in question is obtained as
the algebra of the differential operators
$X_a=e^\alpha_a\partial_\alpha$.}. $\eta_{ab}(t)$ are general
time-dependent coefficients which can be solved for from the Einstein
equations, with components decomposed along the frame. Assuming a
spatially homogenous dilaton gives $\del_a\Phi=e_a^\al\del_\al\Phi=0$,
with $\del_0\Phi$ nonvanishing, so that $R^a_{(a)}$ vanish, with
$R^0_0={1\over 2} (\del_0\Phi)^2$. More details on the Bianchi IX
solution can be found in Appendix F.

The main point of BKL is that close to a spacelike singularity, physics
becomes ultralocal. For dilaton driven cosmologies, this results in a
Kasner-like solution in which the time dependent part of the dilaton
is precisely of the form we have been analyzing.

A simple illustration is provided by a general conformally flat
boundary metric and dilaton
\be
ds^2={F(\bar{x},t)}\,\left(-dt^2+d\bar{x}^2
\right)\hspace{.5in} \Phi=\Phi(\bar{x},t). \label{conf-fields}
\ee
Near a singularity (which may be chosen to be at $t=0$ without loss of
generality) we will assume that space derivatives may be ignored compared to
time derivatives (which would typically blow up). However terms which
contain mixed derivatives need to be retained \cite{Landau}. This
results in the following system of equations for $\Phi$ and
$f \equiv \log F$
\bea  {\partial_t}^2 f(\bar{x},t)+  [{\partial_t} f(\bar{x},t)]^2&=&0
\nonumber\\
{\partial_t}f(\bar{x},t)  {\partial_t} \Phi(\bar{x},t)+
{\partial_t}^2
\Phi(\bar{x},t)&=&0\nonumber\\
3\,{\partial_t}^2f(\bar{x},t)+[{\partial_t}\Phi(\bar{x},t)]^2&=&0 \nn \\
{\partial_t}f(\bar{x},t){\partial_i}f(\bar{x},t)-2\, \
{\partial_t}{\partial_i}f
(\bar{x},t)-{\partial_t}\Phi(\bar{x},t){\partial_i}\Phi(\bar{x},t)
& = & 0 .
\label{1-order}
\eea
The general solution of (\ref{1-order})is
given by
\be
\Phi(\bar{x},t)= \sqrt{3}\, \ln{(t+D(\bar{x}))}
+C_1(\bar{x}),\hspace{.5in}f(\bar{x},t)=
\ln{(t+D(\bar{x}))}+C_2(\bar{x}).
\ee
The last equation in
(\ref{1-order}) imposes the following relation on $C_2$ and $C_1$,
\be
C_2(\bar{x})=\sqrt{3}\, C_1(\bar{x}).
\ee
By choosing
$D=0$, one can see that the behavior of the fields is the same as
in eq.(\ref{metric1sol2}) and eq.(\ref{dil1sol2}) for the symmetric Kasner case.

A similar result holds for a general class of diagonal metrics with
non-vanishing Weyl tensor \cite{BK}.

In fact, in dilaton driven cosmology for any space-time dimension
greater than 3, the approach to a singularity is characterized by a
{\em finite} number of oscillations between Kasner-like solutions.
Consider for example homogeneous cosmologies of type Bianchi IX
\be
ds^2 = -dt^2 + (a_1^2(t) l_\alpha l_\beta +
a_2^2(t) m_\alpha m_\beta + a_3^2(t) n_\alpha n_\beta)\ dx^\alpha
dx^\beta,
\ee
where $l, m, n$ are the three frame vectors
$e^1,e^2,e^3$. The Kasner-like solutions are obtained when the spatial
curvatures can be ignored,
\ben
a_i (t) \sim t^{p_i}~~~~~~~~~~, \Phi \sim \alpha \log (t)
\een
where the Kasner exponents satisfy
\ben
\sum_i p_i =1\ , ~~~~~~~~~~\sum_i p_i^2 = 1 - \frac{\alpha^2}{2}.
\een
The effects of the spatial curvature results in oscillations between
different sets of $p_i$'s till all the $p_i$'s are positive.

The transition between different Kasner regimes lead to an interesting
attractor behavior. Consider the case of 3+1 dimensions
Let $p_-$ denote a negative Kasner exponent and $p_+>0$
being either of the other two positive exponents. These transitions
can be expressed as the iterative map
\be
p_i^{(n+1)} = {-p_-^{(n)}\over 1+2p_-^{(n)}}\ , \quad p_j^{(n+1)} =
{p_+^{(n)}+2p_-^{(n)}\over 1+2p_-^{(n)}}\ , \quad \al_{(n+1)} =
{\al_n\over 1+2p_-^{(n)}}\ ,
\ee
for the bounce from the $(n)$-th to the $(n+1)$-th Kasner regime
with exponents $p_i,p_j$. With each bounce, we see that $\al$
increases, shrinking the allowed space of $\{p_i\}$, as detailed
in Appendix \ref{sec:unibehav}. For $\al\geq 1$, the allowed window of $\{p_i\}$ pinches
sufficiently forcing all $p_i>0$, at which point oscillations cease
and the system settles in the $p_i>0$ attractor region. The rate of
increase of $\al$ is small for small $\al$, since\
$\al_{n+1}-\al_n = \al_n ({-2p_-\over 1+2p_-})$, so that a system
with near constant dilaton ($\al\sim 0$) takes a long iteration time
to ``flow'' towards the $p_i>0$ attractor region.

Furthermore many distinct initial Kasner regimes can flow to the same
attractor point characterized by a set of positive $p_i$'s. The
details of the derivation are given in Appendix \ref{sec:unibehav}. Note that in the
absence of a dilaton the number of such oscillations is infinite even
though the proper time to the singularity is finite \cite{Landau,
misnerBKL,BK}, simply because in this case all the $p_i$'s cannot be
positive.

Finally, these attractor flows exhibit some degree of chaotic
behavior, in the sense that small changes to the initial conditions
give rise to drastic changes in the final endpoints, as elaborated
in Appendix \ref{sec:unibehav}. A quick glimpse at this is obtained by considering
exponents $\{p_1,p_2,p_3,\al=0\}$ corresponding to a non-dilatonic
asymmetric Kasner cosmology which oscillates indefinitely, and
perturbing infinitesimally. Now $\al^2=2(1-\sum {p_i'}^2)$ is
generically nonzero (although small). This latter set
$\{p_i',\al\neq 0\}$ thus flows to the attractor region, while the
former oscillates indefinitely, showing that a small change in the
former gives a drastically different endpoint.

Note that in our setting we have frozen the 5 form field strength and
all the other supergravity fields. This is because we want to embed
BKL type cosmologies in the AdS setup used in this paper. In general
supergravity theories, a BKL type analysis shows that a general
solution (which excites all fields) in the supergravities which follow
from string theories and 11d supergravity exhibit an infinite number
of oscillations between different Kasner regimes \cite{Damour:2002et,
DH:2000}, similar to pure gravity \cite{Landau,
misnerBKL,BK}.

The fact that the general BKL analysis for gravity-dilaton system can
carried over to a discussion of a class of AdS cosmologies is
interesting.  However, the symmetric Kasner is the only solution
whose Weyl curvature vanishes \footnote{Analysing the Weyl tensor components shows that the Weyl tensor
vanishes identically only for flat space and the symmetric Kasner
spacetime. For a generic asymmetric Kasner spacetime with exponents
$(p_1,p_2,p_3)$, some of the nonvanishing Weyl tensor components
diverge as $t^{2p_m-2}$, where $p_m$ is one of the exponents $p_i$.}.
 In the AdS context this means that it
is only for this case that we can have PBH transformations to choose a
flat boundary. In other cases, the metric on which the dual CFT lives
is generically singular and one would expect that the gauge theory
would be singular as well.

\appendix
\section{Dilaton Couplings in the Yang-Mills Theory}\label{sec:dilcoup}

In this appendix we discuss the Lagrangian which appears in eq.(\ref{lagn4}),
especially the dependence of the  dilaton in it.

In the standard AdS/CFT dictionary the operator dual to the dilaton is determined by
the superconformal symmetry to be an appropriate descendant of the chiral primary obtained by symmetrizing
two scalars.
Of particular importance to  this paper is the fact that in eq.(\ref{lagn4}) the
dilaton only couples to the kinetic energy term for the gauge fields and does not couple to the
kinetic energy
terms of the scalars and the fermions.
 To see this it is enough to consider the $U(1)$ theory.
We follow the notation in \cite{keni}.
The supercharges are $Q^I_\alpha, \bar{Q}_{I\dot{\alpha}}$.
Here, the index $I$ upstairs (downstairs)  denotes a $4$ ($\bar{4}$) of $SU(4)$,
and $\alpha, \dot{\alpha}$ are indices for the two different spinor representations of $SO(3,1)$.

The scalars transform like a real $6$-dimensional representation  of $SU(4)$.  For the limited purpose
of carrying out the supersymmetry analysis it is useful to denote the scalar fields by,
$\Phi_{[IJ]}$, where the square brackets indicated antisymmetrization.  The scalars satisfy the condition,
\be
\label{condr}
(\Phi_{[IJ]})^*={1\over 2}\epsilon^{IJKL}\Phi_{KL}.
\ee
The traceless symmetric product of two scalars gives rise to a field which transforms in the $20$ dimensional
representation of $SU(4)$. This is a chiral primary of the full superconformal algebra.
We denote it by,
\be
\label{trs}
T_{IJKL}=\Phi_{[IJ]}\Phi_{[KL]} -{1\over 4!}\epsilon_{IJKL}\Phi_{[PQ]}\Phi_{[RS]} \epsilon^{PQRS}.
\ee
Note this field satisfies the tracelessness condition, $\epsilon^{IJKL}T_{IJKL}=0$.

The operator which the complexified dilaton-axion couples to is, \cite{keni},
\be
\label{opcd}
\hat{O}=\epsilon^{\alpha \beta}\epsilon^{\gamma \delta} Q^I_\alpha Q^J_\beta Q^K_\gamma Q^L_\delta T_{IJKL}.
\ee

The supersymmetry transformations in the $U(1)$ theory are (upto possible numerical factors):
\beqa
\label{sustra}
[Q_\alpha^I,\Phi_{JK}]& = & \delta^I_J\Psi_{k\alpha}-\delta ^I_K \Psi_{J\alpha}  \\
\{Q^I_\beta, \Psi_{I\alpha} \} & = & \delta^I_J F_{\alpha \beta},
\eeqa
and,
\be
\label{last}
[Q^I_\alpha, F_{\beta\gamma}]=0.
\ee

Here $\Psi_{I\alpha}, F_{\alpha, \beta}$ denote the fermionic partners and the gauge fields.
It is then easy to see that this gives
\be
\label{opdd}
\hat{O}\sim F^2.
\ee
In particular $\hat{O}$ does not contain any  coupling  to the  scalar or fermion kinetic terms.

This result is consistent with eq.(\ref{lagn4}). It is also consistent with the statement that the dilaton couples
to the on-shell Lagrangian once we allow for a total derivative term involving the scalars.

A further check on eq.(\ref{lagn4}) may be obtained as follows. Consider deformed $AdS_5
\times S^5$ with a ten dimensional string frame metric
$g_{\alpha\beta} (x)$ and a dilaton $\Phi (x)$. The indices
$(\alpha,\beta)$ are 10 dimensional indices, e.g. $\alpha = (\mu,a)$
where $\mu = 0,\cdots 3$ and $a = 1 \cdots 6$. We will only consider
backgrounds such that $g_{a\mu} = 0$.
Then the Lagrangian for the
dual theory for small deformations is,
\bea
{\cal L} & = & {\sqrt{-g_4}}~e^{-\Phi} {\rm Tr}~
[ -\frac{1}{4}g^{\mu \mu^\prime}g^{\nu
  \nu^\prime}
F_{\mu\nu}F_{\mu^\prime \nu^\prime} \nn \\
& & - \frac{1}{2} g^{\mu\nu}D_\mu X^a D_\nu X^b~g_{ab}
+ \frac{1}{4} [ X^a , X^b ][X^c , X^d ] g_{ac} g_{bd} \nn \\
& & +\frac{1}{2} {\bar{\Psi}}\Gamma^A~e^\mu_A~[-iD_\mu , \Psi ] +
\frac{1}{2} {\bar{\Psi}} \Gamma^A~e^a_A~[ X^b , \Psi ] g_{ab} ].
\label{ymstring}
\eea
Here $g_4$ denotes the determinant ${\rm det}(g_{\mu\nu})$ and
$A$ denotes a frame index and $e^a_A$ is the string frame vierbein.
One way to see this is to consider the Yang-Mills theory in the
Coulomb branch with $SU(N) \rightarrow SU(N-1) \times U(1)$. Then the
effective action for the $U(1)$ part should be given by the DBI action
for a 3-brane in this geometry. In this action, the dilaton factor
$e^\Phi$ appears as an overall factor, provided everything is
written in terms of the string frame metric. The leading order (two
derivative) terms of this action can be obtained by simply replacing
the $SU(N)$ fields in the original YM action by the $U(1)$ part. It is
then easy to see that if we make this replacement in (\ref{ymstring}),
we get the correct leading terms of the DBI action.

We now need to express the Lagrangian (\ref{ymstring}) in terms of the
ten dimensional Einstein frame metric $G_{\alpha\beta}$
\ben
G_{\alpha\beta} = e^{-\Phi/2}~g_{\alpha\beta}.
\een
This leads to the Lagrangian
\bea
{\cal L} & = & {\sqrt{-G}}~ {\rm Tr}~
[ -\frac{1}{4} e^{-\Phi}~G^{\mu \mu^\prime}G^{\nu
  \nu^\prime}
F_{\mu\nu}F_{\mu^\prime \nu^\prime} \nn \\
& & - \frac{1}{2} G^{\mu\nu}D_\mu X^a D_\nu X^b~G_{ab}
+ \frac{1}{4} e^\Phi [ X^a , X^b ][X^c , X^d ] G_{ac} G_{bd} \nn \\
& & +\frac{1}{2} e^{\Phi/4}
{\bar{\Psi}}\Gamma^A~(e_E)^\mu_A~[-iD_\mu , \Psi ] +
\frac{1}{2} e^{3\Phi/4}
{\bar{\Psi}} \Gamma^A~(e_E)^a_A~[ X^b , \Psi ] G_{ab} ],
\label{ymeinstein}
\eea
where $(e_E)^a_A$ denotes the Einstein frame vierbein. Now consider
the field redefinition for the fermion fields
\ben
\Psi \rightarrow e^{-\Phi/8} \Psi.
\een
This will absorb the dilaton factor in front of the quadratic term,
but will give rise to an additional term of the form \footnote{This includes the transformation
of the spin connection.},
$  {\bar{\Psi}}\Gamma^A \Psi (\partial_\mu \phi) (e_E)^\mu_A$.
However, ${\bar{\Psi}}\Gamma^A \Psi = 0$ by virtue of the Majorana
condition.

In our setup, it is the Einstein metric which is flat. This gives rise
to the equation (\ref{lagn4}).

Finally, let us mention that eq.(\ref{lagn4}) is invariant under the conformal transformation,
$g_{\mu\nu}\rightarrow \Omega^2 g_{\mu\nu}$, with the scalars and fermions
transforming in the standard manner,
once the ${1\over 6} R X^2$ term is also  included.

\section{Reduction to Single Harmonic Oscillator and \\
The Schrodinger picture wave function} \label{sec:shm}

Consider the $3+1$ dimensional quadratic theory,
\be
\label{q4dim}
S=\int dt d^3x {1\over 2} [\dot{X}^2 -(\partial_i X)^2 -m^2(t) X^2].
\ee
We work in a box of volume $V \equiv L^3$, with periodic boundary conditions.
Define the modes, $X_{\bf{n}}$ to satisfy the equation,
\be
\label{exp}
X=\sum_{\bf{n}} X_{\bf{n}} e^{i[{2 \pi\bf{n}\cdot x \over L}] }.
\ee
The action, eq.(\ref{q4dim}), becomes,
\be
\label{q2}
S=\int dt {V\over 2} [\dot{X}_{\bf{n}} \dot{X}_{-\bf{n}} + ({2\pi \bf{n} \over L})^2 +m^2(t)|X_{\bf{n}}|^2].
\ee
Taking
\be
\label{redx}
\sqrt{V} X_{\bf{n}}\rightarrow  X_{\bf{n}}
\ee
 gives the action for a single mode, 
 \be
\label{lagsm}
S=\int dt[|{\dot X}_{\bf{n}}|^2 -\omega^2(t) |X_{\bf{n}}|^2],
\ee
with a time dependent frequency
\be
\label{defw}
\omega^2(t)=w^2(t)=({2\pi \bf{n} \over L})^2+m^2(t).
\ee

Next,  we calculate the wave function for the ground state, eq.(\ref{gndstate}),
in the $X$, and $\tilde{X}$ descriptions.

In the position space representation the ground state wave function is given by,
\be
\label{gndswfa}
\psi(x,t)=<x,t|0>,
\ee
where $|x,t>$ is an eigenstate of the  operator $\hat{X}(t)$,  eq.(\ref{hatx}).
By definition,  $|x,t>$ satisfies the condition,
\be
\label{condxt}
\hat{X}|x,t>=x|x,t>.
\ee
In this representation, $\hat{P}$ the canonically conjugate variable to $\hat{X}$, is the operator,
\be
\label{opp}
\hat{P}=-i\partial_x.
\ee

Now from the definition of the ground state, eq.(\ref{gndstate}), and the expression for
$\hat{X}, \hat{P}$, in terms of the creation and annihilation operators, eq.(\ref{hatx}),
eq.(\ref{defp}), it  follows
that the ground state satisfies the condition,
\be
\label{condgsa}
\hat{P}|0>=({\dot{f} \over f})^*\hat{X}|0>.
\ee
From the properties discussed above it then follows  that the wave function
$\psi(x,t)$, eq.(\ref{gndswfa}),  satisfies  the equation,
\be
\label{prognsb}
-i\partial_x \psi(x,t)=({\dot{f} \over f})^*x\psi(x,t).
\ee
This can be easily integrated to give,
\be
\label{prognsc}
\psi(x,t)=C(t) e^{i[({\dot{f} \over f})^*{x^2 \over 2}]}.
\ee
The time dependent function $C(t)$ is determined by requiring that Schrodinger's equation,
\be
\label{segs}
-{1\over 2} \partial_x^2\psi(x,t) + {1\over 2} \omega^2(t) \psi(x,t) = i \partial_t \psi(x,t),
\ee
 is met.
It is straightforward to see that this gives, eq.(\ref{gndspsi}).

The function,  $f(t)$,  is defined in eq.(\ref{deff}). At small $t$,
\be
\label{hsmt}
H_\nu^{(1)}(-\omega_0 t) \simeq i N_\nu(-\omega_0 t) \sim c_1 (\omega_0 t)^{-\nu},
\ee
where $c_1$ is a constant.
 This leads to eq.(\ref{bform}).

The probability density to find the system between $x$ and $x+dx$ is given by $|\psi(x,t)|^2$.
From eq.(\ref{gndspsi}) this  takes the form,
\be
\label{pgs1}
|\psi(x,t)|^2={|A|^2 \over |f|}e^{-[{\omega_0 x^2 \over |f|^2}]}.
\ee
We have used the fact that $f(t)$ solves  eq.(\ref{eqf}), and has the asymptotic value,
eq.(\ref{asf}). This means that  the Wronskian, which is time independent, is given by,
\be
\label{wrons}
f(t) \dot{f}^*(t)-f^*(t) \dot{f}(t)=2i \omega_0.
\ee

We see from eq.(\ref{pgs1}) that the probability density is a Gaussian with a width,
$\sqrt{(\Delta x)^2} \sim |f|$. This diverges as $t\rightarrow 0$, since $|f|$ blows up,
eq.(\ref{bform}).  Thus the probability density gets more and more uniformly spread out as one approaches the singularity.

The phase factor, eq.(\ref{bhnsing}), arises from the limiting form of $f(t)$ given
 in eq.(\ref{bform}).

Next we turn to the wave function for the ground state in the $\tilde{X}$ variable.
The steps are analogous to those above. The $\tilde{X}$ and $X$ variables are related by,
eq.(\ref{nvarx}).
From the Lagrangian, eq.(\ref{lagsmt}),
 it follows that the conjugate momentum, $\hat{\tilde{P}}$, is given by,
\be
\label{hattildep}
\hat{\tilde{P}}=e^{-\Phi}\dot{\tilde{X}}=e^{-\Phi/2}(\hat{P}+{\dot{\Phi} \over 2} \hat{X}),
\ee
where we have used the relation, $\dot{\hat{X}}=\hat{P}$.

The relation, eq.(\ref{condgsa}), then leads to,
\be
\label{relb}
\hat{\tilde{P}}|0>=e^{-\Phi}(({\dot{f}\over f})^* + {\dot{\Phi}\over 2})\hat{\tilde{X}} |0>.
\ee
Let $|\tilde{x},t>$ be  eigenstates of $\tilde{X}$, satisfying the condition \footnote{We also
require that the completeness relation,
$$\int d\tilde{x} |\tilde{x},t><\tilde{x},t|=\bf{I}$$
is satisfied. A similar relation, with $\tilde{x}$, replaced by $x$, is also satisfied
 by the states, $|x,t>$. This tells us that  the states $|\tilde{x},t>$ are related to the states,
$|x,t>$ introduced above, by the relation,
$$|\tilde{x},t>=e^{-\Phi/4} |x> = e^{-\Phi/2}| \tilde{x}, t>.$$}.
\be
\label{est}
\hat{\tilde{X}}|\tilde{x},t> =\tilde{x}|\tilde{x},t>.
\ee
The wave function in the $|\tilde{x}>$ representation, $\tilde{\psi}(\tilde{x},t)$, then satisfies
the condition,
\be
\label{relc}
-i\partial_{\tilde{x}}\tilde{\psi}=e^{-\Phi}[({\dot{f} \over f})^*+{\dot{\Phi}\over 2}]\tilde{x}
\tilde{\psi}.
\ee
In addition the Schrodinger equation which takes the form,
\be
\label{set}
-{e^{\Phi} \over 2} {\partial^2 \tilde{\psi} \over \partial \tilde{x}^2}+ {e^{-\Phi} \over 2}
\omega_0^2 \tilde{\psi}=i\partial_t \tilde{\psi},
\ee
must be satisfied. This leads to the solution, eq.(\ref{gndstilde}).
The probability density $|\tilde{\psi}|^2$, eq.(\ref{probt}), is then obtained as in the discussion
  above leading upto, eq.(\ref{pgs1}).  To understand the behavior of the phase factor discussed
in eq.(\ref{phase}), we note that  $f(t)$  is defined in eq.(\ref{deff}). At small $t$ it then follows
from the behavior of the Neumann function $N_\nu$, that,
\be
\label{smalltf}
f(t)\simeq c_1 (-\omega_0 t)^{{1\over 2}-\nu}(1+ c_2 (-\omega_0 t)^2).
\ee
Thus,
\be
\label{valp}
({\dot{f}\over f})^*+{\dot{\Phi}\over 2} ={1\over t}({1 -2 \nu+p \over 2}) + 2 c_2 \omega_0^2 t.
\ee
Here we have used, eq.(\ref{gendil}).
This leads to eq.(\ref{phase}), after noting, eq.(\ref{defnu}).

The expectation value for $\hat{\tilde{P}}^2$ can be calculated from the wave function eq.(\ref{gndstilde}).
Alternatively, for this purpose, we can directly work in the Heisenberg picture.
From eq.(\ref{hattildep}), and the expression for $\hat{X}, \hat{P},$ in term of the creation and
annihilation operators, eq.(\ref{hatx}), eq.(\ref{defp}),  we get that in the vacuum state,
\be
\label{p2vac}
<\hat{\tilde{P}}^2>={e^{-\Phi} \over 2 \omega_0 }|\dot{f}+{\dot{\Phi}\over 2} f|^2 .
\ee
Using, eq.(\ref{valp}) and related discussion above, this leads to eq.(\ref{phat}).

A similar analysis can be carried out for a coherent state, defined by,
\be
\label{defcoh}
a|s>=\alpha |s>.
\ee
This leads to a wave function,
\be
\label{wfcoherent}
\tilde{\psi}(\tilde{x},t)={A \over \sqrt{f^* e^{\Phi/2}}}
 e^{[ie^{-\Phi}{\tilde{x}^2\over 2}[({\dot{f}\over f})^*+{\dot{\Phi}\over 2}]}
e^{[{\alpha \sqrt{2\omega_0}\tilde{x} \over f^*e^{\Phi/2}}]}
e^{[i\omega_0\alpha^2 \int {dt \over (f^*)^2}]} .
\ee
The extra terms, compared to the ground state wave function, which are dependent on $\alpha$, are
both well defined in the limit $t\rightarrow 0$. Thus this wave function has the same  type
of singularity
 as the ground state wave function.


\section{Subleading Contributions to Energy}\label{sec:subenergy}

In this appendix we calculate the subleading contributions to the energy. These contributions
would be the dominant ones if $<V>$ vanishes, as discussed in section \ref{sec:enblow},
 and need to be calculated to understand when the
energy remains finite, as $t\rightarrow 0$.

A subleading contribution arises from the kinetic energy term.
From the wave function, eq.(\ref{sol}), we find that for $p>1$ this is given by,
\be
\label{exke}
<KE> \simeq {e^\Phi\over 2} G^2 <(V')^2> \sim (-t)^{(2-p)},
\ee
and diverges if $p>2$.
For $p<1$, since $G(t)$ is small near $t=0$,
\be
\label{exket}
<KE>\simeq {e^{\Phi}\over 2}\int dx|\psi_0'|^2 \sim (-t)^p.
\ee
This does not diverge as $t\rightarrow 0$.

Another subleading correction arises due to a correction in the absolute magnitude of $\psi$
which in turn leads to a correction in $<V>$.
We write,
\be
\label{expw}
\psi(x,t)=e^{-iG(t)V(\tilde{x})} \psi_0(\tilde{x})[1+S_1(\tilde{x},t)].
\ee
Since we are interested in the corrections to the
absolute value of $\psi$ we take $S_1$ to be real.
From the Schrodinger equation we get,
\be
\label{sethree}
{e^\Phi \over 2} [-Im ({\psi_0{''} \over \psi_0})+2G(t)V'Re({\psi_0'\over \psi_0}) + G(t) V''] =
{\partial S_1 \over \partial t}.
\ee
For $p>1$ the second and third  terms within the square brackets on the left hand side dominate,
leading to,
\begin{eqnarray}
\label{exs1a}
S_1 & = & [\int dt e^\Phi G(t)] [ V'Re({\psi_0' \over \psi_0})+{V^{''} \over 2}] \\
    &=& {1\over 2 (1-p)} t^2 [V'Re({\psi_0' \over \psi_0})+{V^{''} \over 2}].
\end{eqnarray}
We see that this goes like $t^2$, as $t\rightarrow 0$ and does not diverge.

For $p<1$ the first term on the left hand side of eq.(\ref{sethree}) dominates, giving,
\begin{eqnarray}
\label{exs1b}
S_1& = & {1\over 2} Im({\psi_0^{''}\over \psi_0})[\int dt e^\Phi] \\
   &=& -{1\over 2(1+p)} Im({\psi_0^{''}\over \psi_0})   (-t)^{(1+p)}.
\end{eqnarray}
This term goes like $(-t)^{(1+p)}$ and also does not diverge.
Since $S_1$ vanishes as $t\rightarrow 0$, the resulting correction to $<V>$ and therefore to
 the energy also vanishes.

Thus  the conclusion is that that except for the case where $p>2$, in which case the kinetic
energy itself gives a divergent contribution,
it is enough to have $<V>$ as defined in eq.(\ref{exv}) to
vanish,to ensure that the expectation value of the energy stays finite.

\section{Particle Production}\label{sec:partprod}

In this appendix we detail the calculation of particle production at
the quadratic level in the case where $e^\Phi$ does not become zero at
any point, but can become small. For this purpose we choose a dilaton
profile of the following form
\bea
e^{\Phi (t)} & = & g_s |t|^p~~~~~~~~~~~~~~~~|t| > \epsilon \nn \\
e^{\Phi (t)} & = & g_s |\epsilon|^p~~~~~~~~~~~~~~~~|t| < \epsilon .
\label{ppone}
\eea
We will first perform the analysis for each individual momentum mode,
$\tX_k$.

 As explained above, one should work in the variables $\tX$ since
 these are the variables which have a finite limit as $t \rightarrow
 0$. The equation of motion for $\tX_k$ is
\ben
\left[ \frac{d}{dt} \left( e^{-\Phi (t)}\frac{d}{dt} \right) +
\omega_0^2~e^{-\Phi
 (t)} \right] \tX_k = 0 .
\label{pptwo}
\een
Clearly, it is convenient to work with a time variable $\tau$ defined by
\ben
e^{-\Phi (t)}\frac{d}{dt} = \frac{d}{d \tau}.
\label{ppthree}
\een
As is standard, we will solve (\ref{pptwo}) separately in the regions
$ t < -\epsilon$, $-\epsilon < t < \epsilon$ and $t > \epsilon$ and
then match $\tX_k$ and $\partial_\tau \tX_k$ across $t = \pm
\epsilon$. With the profile given in (\ref{ppone}), a solution which
is purely positive frequency at $t \rightarrow -\infty$ is given by
\bea
\tX_{k} (t) & = & (-\omega_0 t)^\nu~H_\nu^{(1)}(-\omega_0 t) \quad  \quad \quad
\quad \quad \quad
t \leq
-\epsilon \nn \\
\tX_{k} (t) & = & A~{\rm exp} \left[ i \frac{\omega_0 t^{p+1}}{\epsilon^p
    (p+1)} \right] + B~{\rm exp} \left[ - i \frac{\omega_0 t^{p+1}}{\epsilon^p
    (p+1)} \right] \quad \quad  -\epsilon \leq t \leq \epsilon \nn \\
\tX_{k} (t) & = & (\omega_0 t)^\nu \left[ C~H^{(1)}_\nu (\omega_0 t)
+ D~H^{(2)}_\nu (\omega_0 t) \right] \quad \quad t \geq \epsilon ,
\label{ppfour}
\eea
where $\nu$ has been defined in (\ref{defnu}).
The Bogoliubov coefficients $C$ and $D$ will be determined by the
matching conditions.
After a standard calculation we get the following expressions for $C$
and $D$ :
\bea
C & = & \frac{i\pi}{4} (\omega_0 \epsilon)
\{ \cos \left( \frac{2\omega_0 \epsilon}{p+1}
\right)  \left[
  H^{(1)}_\nu (\omega_0 \epsilon) H^{(2)}_{\nu -1}(\omega_0\epsilon) +
H^{(2)}_\nu (\omega_0 \epsilon) H^{(1)}_{\nu -1}(\omega_0\epsilon)
\right]
\nn \\
& - & \sin \left( \frac{2\omega_0 \epsilon}{p+1} \right) \left[
  H^{(1)}_{\nu -1} (\omega_0 \epsilon) H^{(2)}_{\nu -1}(\omega_0\epsilon) -
H^{(2)}_\nu (\omega_0 \epsilon) H^{(1)}_{\nu}(\omega_0\epsilon) \right] \}
\nn \\
D & = & - \frac{i\pi}{4} (\omega_0 \epsilon) \{ 2 \cos \left( \frac{2\omega_0
  \epsilon}{p+1}  \right)  H^{(1)}_\nu (\omega_0 \epsilon) H^{(1)}_{\nu
  -1}(\omega_0\epsilon) \nn \\
& - & \sin \left( \frac{2\omega_0 \epsilon}{p+1} \right) \left[
(H^{(1)}_{\nu -1} (\omega_0 \epsilon) )^2 -
(H^{(1)}_{\nu} (\omega_0 \epsilon) )^2 \right]
\} .
\label{ppfive}
\eea
In deriving these we have used the following property for any Bessel
function $Z_\nu (x)$:
\ben
\frac{d}{dx} [ x^\nu Z_\nu (x) ] = x^\nu Z_{\nu-1} (x).
\een
A straightforward calculation verifies the unitarity relation
\ben
|C|^2 - |D|^2 = -1 .
\een

Let us first consider the limit $\omega_0 \epsilon \ll 1$.
Using the standard expansions for the Hankel functions,
\bea
H^{(1)}_\nu (x) = \frac{i x^\nu}{2^\nu \sin(\pi\nu)\Gamma (1+\nu)}  & \{
& e^{-i\pi\nu} \left( 1 - \frac{x^2}{4(\nu+1)} + O(x^4)  \right) \nn
\\
& - &
\left( \frac{x}{2}
\right)^{2\nu} \frac{\Gamma(1+\nu)}{\Gamma (1-\nu)}
\left( 1-\frac{x^2}{4(1-\nu)} + O(x^4) \right) \},
\eea
we find,
\ben
C = \frac{i\pi}{4}\frac{1}{\sin^2 (\pi\nu)~2^{2\nu}~[\Gamma(1-\nu)]^2}
\left[ -2^{2p+2} \frac{(\omega_0 \epsilon)^{1-p}}{1-\nu} -
  2^{p+2}\nu~\cos(\pi\nu)~ \frac{\Gamma(1-\nu)}{\Gamma(1+\nu)} +
  \frac{2(\omega_0 \epsilon)^{1-p}}{p+1} 2^{2p+2} \right].
\een
Thus there is a qualitatively different behavior of the Bogoliubov
coefficient for $p > 1$ and for $p < 1$ as $(\omega_0 \epsilon)
\rightarrow 0$. When $p > 1$ the coefficients $C$ and $D$ both diverge
in this limit (of course maintaining the unitarity relation). When $p <
1$ they both tend to finite limits
\ben
{\rm Lim}_{\omega_0 \epsilon \rightarrow 0}~C = -i \cot(\pi
\nu)~~~~~~~~~~~~~~~~~ {\rm Lim}_{\omega_0 \epsilon \rightarrow 0}~D =
 -i  e^{-i\pi\nu}{\rm cosec}(\pi\nu) .
\een
This difference between the cases $p < 1$ and $p > 1$
is the Heisenberg picture manifestation of the behavior of the
Schrodinger picture wavefunctional.

The analysis performed above was with an abrupt modification of the
dilaton profile. However we expect that the $\omega_0 \epsilon \ll 1$
behavior would continue to be similar for a smooth modification.

In the above analysis
there is a finite amount of particle production
for every momentum mode for $p < 1$,
independent of the value of $\omega_0$, in the
limit $\omega_0 \epsilon \rightarrow 0$. It is interesting to estimate
the total amount of energy produced. However, for this estimate we
need to perform the calculation for a dilaton profile which tends to a
constant at early and late times, in keeping with our overall
scenario. For a smooth dilaton profile, the
ultraviolet behavior ($\omega_0 \epsilon \gg 1$)
is then expected to be exponentially
damped, $|C|^2 \sim e^{-\omega_0 \epsilon}$, so that the total energy
produced is finite.

It should be emphasized again that all the considerations of this
appendix relates to the quadratic approximation. As we have seen this
is {\em not} a good approximation in our problem. An estimate of the
total amount of energy produced in the real problem has to take into
account the effects of interactions which become stronger at later
times. This requires a lot more detailed knowledge of strong coupling
physics.

\section{The Milne Background}\label{sec:milne}

In this appendix we analyze the behavior when the boundary theory lives in Milne space with metric,
eq.(\ref{bmet3}), with a dilaton,
\be
\label{dilap}
e^\Phi=g_s |\tanh (t)|^{\sqrt{3}}.
\ee
The metric eq.(\ref{bmet3}),  upto the overall conformal factor, $e^{-2t}$, is,
\be
\label{metc}
ds^2=-dt^2+{dr^2\over 1+r^2} + r^2 (d\theta^2+\sin^2\theta d\phi^2).
\ee
This is a space of constant negative curvature.
Since the gauge theory is conformally invariant it is equivalent to consider it
 in the background metric, eq.(\ref{metc}), and with dilaton, eq.(\ref{dilap}).

Below we first analyze a scalar field, with Lagrangian \footnote{The conclusions would be
 essentially the same without the curvature coupling term, $ {1\over 12} R \tilde{X}^2$.}
\be
\label{cfs}
S=\int d^4x \sqrt{-g} e^{-\Phi} \{ {1\over 2}(\partial \tilde{X})^2 + {1\over 12} R \tilde{X}^2 \},
\ee
in the background with metric, eq.(\ref{metc}), and dilaton, eq.(\ref{dilap}). Thereafter turn to the gauge field. The analysis in the
 scalar field theory, eq.(\ref{cfs}), is not identical
to the gauge theory, but quite analogous.

We can mode decompose the scalar field into modes which are eigenfunctions of the $3$ dim. spatial
 Laplacian. From \cite{BD}, we see that, for the metric, eq.(\ref{metc}), the modes satisfy the equation,
\be
\label{mode}
\nabla^3 y_{\bf{k}} =-(k^2+1) y_{\bf{k}}.
\ee
The functions, $y_{\bf{k}}$, are normalized to satisfy the condition,
\be
\label{norm}
\int d^3x\sqrt{h} y_{\bf{k}}(x) y_{\bf{k}'}^*(x')=\delta^3(\bf{k}-\bf{k}').
\ee
Here $h_{ij}$ is the spatial part of the metric, eq.(\ref{metc}).
We can expand the field, $\tilde{X}$, in these modes,
\be
\label{xtexp}
\tilde{X}=\int d^3k \tilde{X}_{\bf{k}}(t) y_{\bf{k}}.
\ee
This gives rise to decoupled oscillators for each mode, with the Lagrangian,
\be
\label{deco}
S=\int dt d^3k e^{-\Phi}\{|\dot{\tilde{X}}_{\bf{k}}|^2 -k^2 |\tilde{X}_{\bf{k}}|^2\},
\ee
where we have used the fact that the Ricci scalar, $R=-6$, for the metric. eq.(\ref{metc}).
We see that for each mode the Lagrangian, eq.(\ref{deco}), is essentially the same as eq.(\ref{lagtilde}),
with $k^2$ being identified with $\omega_0^2$. Since the dilaton  asymptotically goes to a
 constant here, eq.(\ref{dilap}), the Lagrangian for each mode reduces to that of a standard
harmonic oscillator in the far past or future.

Our discussion in section 5 then leads to the conclusion that
the  wave function has a  phase factor which  is singular as $t\rightarrow 0$,
in this case as well.  The phase factor is given by, eq.(\ref{wfunctional}). The potential energy,
$V[\tilde{X}]$, for
the Lagrangian eq.(\ref{cfs}) is,
\be
\label{vm}
V[\tilde{X}]=\int d^3x \sqrt{h} [h^{ij} {1\over 2} \partial_i \tilde{X} \partial_j \tilde{X}
+{1\over 12} R \tilde{X}^2].
\ee

For the case of the gauge field, in the background metric, eq.(\ref{metc}),
 with the dilaton, eq.(\ref{dilap}),
an analysis similar to that carried out here  and in  the previous appendix can be done.
Once again choosing Coulomb gauge is convenient. In this gauge one finds that the wave function
has a phase factor which near the singularity takes the form, eq.(\ref{pgauge}),
with,
\be
\label{potm}
V[A_i(x)]={1\over 4}\int d^3x\sqrt{-h}F_{ij}F^{ij}.
\ee
   This phase factor diverges, leading to a singular wave function.

\section{Universal behavior near singularities}\label{sec:unibehav}

In this appendix we discuss some aspects of the universality of Kasner
like behavior near spacelike singularities in the class of models we
consider.  We will consider ten dimensional metrics of the form
(\ref{gensol}) so that the 3+1 dimensional Ricci tensor $R_{\mu\nu}$
and the dilaton $\Phi(x)$ satisfies the equations (\ref{deom}). It is
therefore sufficient to discuss 3+1 dimensional dilaton cosmologies.

Consider an AdS cosmology where the 4-metric is a Bianchi-IX
spacetime. Using a BKL-type argument \cite{Landau,misnerBKL,BK} we
show the spacetime near the singularity has Kasner-like
behavior. Furthermore, the dilaton drives the system towards an
attractor region, where all the exponents are positive $p_i>0$,
through a finite number of Kasner oscillations. This analysis can be
directly extended to all homogeneous spaces with the results that we
either have no oscillations at all or the number of oscillation is
finite. This suggests that the symmetric Kasner singularity is generic
and independent of the spatial 3-geometry, being either flat or any of
Bianchi homogeneous spaces.\\ It is worth mentioning that with no
dilaton, symmetric Kasner solutions do not exist and the canonical BKL
analysis gives an oscillatory approach to the singularity, with
transitions between distinct asymmetric Kasner regimes.

Let us take the 3+1 dimensional boundary metric $\tg_{\mu\nu}(x)$ in
(\ref{gensol}) to be of Bianchi-IX type, which has the following form
\be ds_4^2 = -dt^2 + (a^2(t) l_\alpha l_\beta +
b^2(t) m_\alpha m_\beta + c^2(t) n_\alpha n_\beta)\ dx^\alpha
dx^\beta
\ee
where $l, m, n$ are the three frame vectors
$e^1,e^2,e^3$,\ (for explicit form of the metric see, for
example, \cite{Landau} page 390). The spatial symmetry algebra here
is $SU(2)$. $a,b$, and $c$ are three independent scale
factors\footnote{If we take equal scale factors $a=b=c$, the spatial
metric becomes\ \
$d\sigma^2=(dx_1^2+dx_2^2+dx_3^2+\cos x^1 dx^2dx^3)$, with
constant curvature, $R_{ij}={1\over 2}\gamma_{ij}, R={3\over
2}$.}. \

If we assume that the dilaton is spatially homogeneous, then\
$\del_a\Phi=e_a^\al\del_\al\Phi=0$, with $\del_0\Phi$
nonvanishing. Decomposing the Ricci tensor along the frame, we
then have, \bea R^1_{(1)} &=& {\dot ({\dot a} b c)\over abc} -
{1\over 2(abc)^2}
[(b^2-c^2)^2 - a^4] = 0\ \nonumber\\
R^2_{(2)} &=& {\dot (a {\dot b} c)\over abc} - {1\over 2(abc)^2}
[(a^2-c^2)^2 - b^4] = 0\ \nonumber\\
R^3_{(3)} &=& {\dot (a b {\dot c})\over abc} - {1\over 2(abc)^2}
[(a^2-b^2)^2 - c^4] = 0\label{II} \eea \be R^0_0 = {\ddot{a} \over
a}+{\ddot{b} \over b}+{\ddot{c} \over c}=-{1\over 2} (\dot\Phi)^2,
\ee and the dilaton field equation is given by \be
\ddot\Phi+\dot\Phi\left( {\dot{a} \over a}+{\dot{b} \over
b}+{\dot{c} \over c} \right) =0. \ee

The above system of ODE's is very difficult to solve analytically
but if we ignore curvature terms (i.e., terms in (\ref{II}) with
no time derivatives) one has\ $a=t^{p_1}, b=t^{p_2},
c=t^{p_3},\ e^\Phi=t^\al$,  \be \sum p_i = 1\ ,\qquad \sum p_i^2 = 1 -
{\al^2\ \over 2} , \label{constraints}\ee as an approximate
solution near the singularity at $t=0$.

In the usual BKL analysis, $\alpha=0$, which forces one of the
$p_i$ to be negative. The negative $p_i$ means that time evolution
towards the singularity necessarily makes one of the curvature
terms ( treated as a perturbation to the time-derivative terms )
dominate the others at some point, for instance $a^4\sim
t^{-|p_1|}$ (if $p_1^{(0)}<0$). This forces the metric to evolve
and transit from one Kasner regime to another according to the
following law \bea p_1^{(1)} = {-p_1^{(0)}\over 1+2p_1^{(0)}}\ ,
\quad && p_2^{(1)} = {p_2^{(0)}+2p_1^{(0)}\over 1+2p_1^{(0)}}\ ,
\quad p_3^{(1)} = {p_3^{(0)}+2p_1^{(0)}\over 1+2p_1^{(0)}}\ . \eea
However, with a nontrivial dilaton, we can have one of the
following situations: All $p_i>0$, in which case no transitions
take place since curvature terms (perturbations) die off as we
approach the singularity \eg\ \
$$[(b^2-c^2)^2 - a^4] \sim -a^4 \sim t^{4/3} \ra 0$$
near the singularity $t\ra 0$. Other Ricci components have similar
behavior. This means that the symmetric Kasner case with all
$p_i={1\over 3}$ is stable against these perturbations as we
approach this dilaton-driven symmetric Kasner singularity and there
is no forced transition to a distinct Kasner regime. The other
possibility is that one of the $p_i$s is negative, in this case, we
can have a finite number of oscillatory transitions between different
Kasner regimes. This occurs since with every transition $\alpha$
increases and as it reaches a specific value ($\alpha=1$) all the
$p_i$ become positive (see (\ref{constraints}), and then no further
transitions occur.\\
If we consider any of the other types of Bianchi spaces, we should
replace the curvature terms in (\ref{II}) with that of this Bianchi
space.
But these terms have no time-derivatives so they will not change
the leading behavior of the solution. Furthermore, these
curvature terms either die off as we approach the singularity, in
which case we have no oscillation at all, or one of them ($p_1<0$)
gets larger as we approach the singularity. This leads to a finite
number of oscillations as in Bianchi-IX. Again the dilaton will
drive the system to an attractor region where the oscillation stops.

We will now see that the dilaton in fact drives the system towards
an ``attractor'' region given by $p_1,p_2,p_3>0$, through a finite
number of oscillations. Once the system reaches this region, there
is no further oscillation.

The equations \be \sum_i p_i =1\ , \qquad \sum_i p_i^2 =
1-{\al^2\over 2}\ , \ee for a Kasner-like cosmology with dilaton
$e^\Phi=t^\al$ can be described by the following parametrization
\be p_1=x\ , \quad p_2 = {1-x\over 2} +
{\sqrt{1-\al^2+2x-3x^2}\over 2}\ , \quad p_3 = {1-x\over 2} -
{\sqrt{1-\al^2+2x-3x^2}\over 2}\ , \ee in terms of $p_1,\al$. For
a solution to exist, the radical being positive forces \be
{1-\sqrt{4-3\al^2}\over 3} \leq p_1 \leq {1+\sqrt{4-3\al^2}\over
3}\ . \ee The above range for $p_1$ can be divided into three
regions. In these regions the values of the $p_i$'s are permuted
among each other, as in figure $(1)$. To avoid redundancy one
should constrain $p_1$ to one region which we choose to be
 \be{1-\sqrt{4-3\al^2}\over 3} \leq p_1\leq {2-\sqrt{4-3\al^2}\over 6} . \ee

The square root here being positive implies $\al^2\leq {4\over
3}$, \ie\ $|\al|\leq {2\over \sqrt{3}}\sim 1.1547$. The lower
limit on $p_1$ becomes positive if $4-3\al^2\leq 1$, \ie\
$\al^2\geq 1$. At this point, $x=0,\ \al^2=1$, all $p_i>0$. This
shrinking of the allowed space of $\{p_i\}$ is a key difference
from the case $\al=0$ without dilaton.

Now let us say the system starts with say $p_1^{(0)}=x^{(0)}<0$.
Then there is a transition to a new Kasner regime with $p_i^{(1)}$ and $\alpha^{(1)}$ given by
\bea p_1^{(1)} = {-p_1^{(0)}\over 1+2p_1^{(0)}}\ , \quad &&
p_2^{(1)} = {p_2^{(0)}+2p_1^{(0)}\over 1+2p_1^{(0)}}\ , \quad
p_3^{(1)} = {p_3^{(0)}+2p_1^{(0)}\over 1+2p_1^{(0)}}\ , \nonumber\\
&& \al^{(1)} = {\al^{(0)}\over 1+2p_1^{(0)}}\ . \eea Now, $p_1^{(0)}<0$
means that $\al^{(1)}>\al^{(0)}$, \ie\ $\al$ increases under the
Kasner transition. More generally, for $p_-<0$ and $p_+>0$ being
either of the other two positive exponents, this can be expressed
as the iterative map
\be\label{p+p-} p_i^{(n+1)} =
{-p_-^{(n)}\over 1+2p_-^{(n)}}\ , \quad p_j^{(n+1)} =
{p_+^{(n)}+2p_-^{(n)}\over 1+2p_-^{(n)}}\ , \quad \al_{(n+1)} =
{\al_n\over 1+2p_-^{(n)}}\ ,
\ee
for the bounce from the $(n)$-th
to the $(n+1)$-th Kasner regime with exponents $p_i,p_j$. The
fixed point of this transformation is\ $\al=0$, and it is unstable
for $p_-<0$\ \ (an iterative map $x_{n+1}=f(x_n)$ has an unstable
fixed point $x_*=f(x_*)$ if $f'(x_*)>1$).
\begin{figure}
\begin{center}
\epsfig{file = 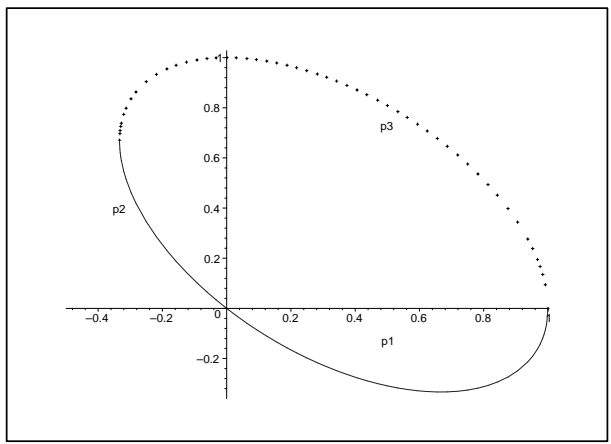, height=8cm,angle=-90,trim=0 0 0
0}\caption{ Here p$_3$ and p$_2$ are plotted as functions of p$_1$
for $\al=0$ case. Notice how the regions $-1/3 \leq p_1 \leq 0$,
\,$0 \leq p_1 \leq 2/3$, \, $2/3 \leq p_1 \leq 1$ have the same
values for $p_i$'s} \label{rootplot}
\end{center}
\end{figure}
Furthermore, the rate of increase of $\al$ is small for small
$\al$, since \be \al_{n+1}-\al_n = \al_n \left({-2p_-\over
1+2p_-}\right) .\ee Thus a system with near constant dilaton
($\al\sim 0$) takes a long iteration time to ``flow'' towards the
$p_i>0$ attractor region (although the flow is ensured due to the
unstable fixed point). For instance, with $p_1^0=x_0=0.3,\
\al_0=0.001$, the system flows (initially slowly) to $p_i>0$ after
15 oscillations, with $\al_{15}=1.0896$.

This flow towards the $p_i>0$ attractor region in $\{p_i\}$-space can
be seen geometrically:\ the intersection of the sphere
$\sum p_i^2=1-{\al^2\over 2}$ with the plane $\sum p_i=1$ is a circle
on the plane. Under the bounce iterations, the sphere radius shrinks and
so the circle radius also shrinks until the circle lies entirely in
the $p_i>0$ quadrant.

The finiteness of the number of oscillations means that the bulk
cosmology flows towards the $p_i>0$ attractor region, driven by
the dilaton.

Before we close this section we would like to comment on the
nature of the flow towards the attractor region. Inverting
(\ref{p+p-}), we get $p_-={p_i^{(n+1)}\over 1+2p_i^{(n+1)}}$ and
so on\footnote{ since $|p_-^{(n)}|<{1\over 2}$ is always true,
therefore $p_i^{(n+1)}={-p_-^{(n)}\over 1+2p_-^{(n)}}>0$. This
means that for each of the other two distinct $p_j^{(n+1)}>0$, we
can potentially trace back to a distinct $p_-^{(n)}<0$. This gives
a tree with two flows starting with a given point
$\{p_i^{(n+1)}\}$. Similarly at every previous point, the tree
forks into two.}. Thus we can trace back from the symmetric
Kasner, giving (upto five iterations) the flow, \be\label{flow35}
(-{1\over 5},{9\over 35},{33\over 35})\ \ra\ (-{5\over 21},{7\over
21},{19\over 21})\ \ra\ (-{3\over 11},{5\over 11},{9\over 11})\
\ra\ (-{1\over 5},{3\over 5},{3\over 5})\ \ra\ ({1\over 3},{1\over
3},{1\over 3})\ .
\ee
An alternative distinct flow to the same symmetric Kasner endpoint begins
at $({-9\over 29},{15\over 29},{23\over 29})$, merging with the above
flow at $(-{3\over 11},{5\over 11},{9\over 11})$. With each step
backwards, $\al$ decreases. This shows that there are multiple
trajectories that get attracted to any of the points in the $\{p_i>0\}$
attractor region, perhaps as for any attractor-like behavior.

Furthermore we suspect that the flow exhibit chaotic behavior,
\ie\ small changes in the initial conditions give rise to
drastic changes in the final endpoints. For example, consider
changing the starting point for the flow (\ref{flow35}) above by a
small perturbation (by ${1\over 70}\sim 0.014$, \ie\ a 7\% change
to the smallest exponent, $-{1\over 5}$). This gives:
\be (-{13\over 70},{9\over 35},{65\over 70})\ \ra\
(-{2\over 11},{13\over 44},{39\over 44})\ \ra\ (-{3\over
28},{2\over 7},{23\over 28})\ \ra\ ({1\over 11},{3\over
22},{17\over 22})\ ,
\ee
the flow endpoint being distinct from the symmetric Kasner.

We have used rational Kasner exponents above for simplicity in
illustration: more generally,
one expects that there exist 'nearby'
(not necessarily rational) Kasner exponents $p_i$ with nonconstant dilaton ($\al\neq 0$) in the
neighborhood of exponents with constant dilaton ($\al=0$). In this
case, a small change in the exponents in the set $\{p_i,\al\neq 0\}$
would give exponents in the set $\{p_i,\al=0\}$, which latter set
belong to the canonical BKL analysis and oscillate forever, thus
exhibiting no attractor behavior. More explicitly consider exponents
$\{p_1,p_2,p_3,\al=0\}$ corresponding to a non-dilatonic asymmetric
Kasner cosmology which oscillates indefinitely, and perturb
infinitesimally as $\{p_1'=p_1+\epsilon,p_2'=p_2,p_3'=p_3-\epsilon\}$.
Now, $\al^2=2(1-\sum {p_i'}^2)\sim 4(p_3-p_1)\epsilon\neq 0$, if
$p_1\neq p_3$. This latter set $\{p_i',\al\neq 0\}$ thus flows to
the attractor region, while the former does not.
These examples and arguments suggest that small perturbations to
initial conditions apparently give rise to large departures from the
endpoints, in other words, chaotic behavior.

\section{Acknowledgments}
We would like to thank Ben Craps, Ian Ellwood, D. Gross, Juan
Maldacena, G. Mandal, Willie Merrell, Shiraz Minwalla, and Spenta
Wadia for discussions and S. Rajeev for a clarification. A.A. and
S.R.D were supported in part by the United States National Science
Foundation Grant Numbers PHY-0244811 and PHY-0555444. S.T. was
supported by a Swarnajayanti Fellowship, DST, Govt. of India.  Some of
the research was carried out during the Monsoon workshop at TIFR. We
thank the organizers, TIFR and ICTS, for support. Most of all we thank
the people of India for generously supporting research in string
theory.

\end{document}